%% file: report.tex
\documentclass[11pt, conference, onecolumn]{headers/IEEEtran}

\usepackage{courier}
\usepackage{setspace}
\input{./headers/psfig}
\input{./headers/mymacros}

\begin{document}

\title{ Data Path Processing in Fast Programmable Routers }

\vspace{1.0in}

\author{
Pradipta De \\
Computer Science Department \\
Stony Brook University \\
Stony Brook, NY 11794-4400 \\
prade@cs.sunysb.edu \\
}
\maketitle
\thispagestyle{empty}

%%%%%%%%%%%%%%%%%%%%%%%%%%%%%%%%%%%%%%%%%%%%%%%%%%%%%%%%%%%%%%%%%%%%%

\begin{abstract}
Internet is growing at a fast pace. The link speeds are
surging toward 40 Gbps with the emergence of faster link
technologies. New applications are coming up which require
intelligent processing at the intermediate routers. Switches and
routers are becoming the bottlenecks in fast communication. On
one hand faster links deliver more packets every second and on
the other hand intelligent processing consumes more CPU cycles
at the router. The conflicting goals of providing faster but
computationally expensive processing call for new approaches
in designing routers.

This survey takes a look at the core functionalities, like
packet classification, buffer memory management, switch
scheduling and output link scheduling performed by a router
in its data path processing and discusses the algorithms that
aim to reduce the performance bound for these operations. An
important requirement for the routers is to provide Quality of
Service guarantees. We propose an algorithm to guarantee QoS
in Input Queued Routers.  The hardware solution to speed up
router operation was Application Specific Integrated Circuits
(ASICs). But the inherent inflexibility of the method is a
demerit as network standards and application requirements
are constantly evolving, which seek a faster turnaround time
to keep up with the changes. The promise of Network Processors
(NP) is the flexibility of general-purpose processors together
with the speed of ASICs. We will study the architectural choices
for the design of Network Processors and focus on some of the
commercially available NPs. There is a plethora of NP vendors
in the market. The discussion on the NP benchmarks sets the
normalizing platform to evaluate these NPs.

\end{abstract}

%%%%%%%%%%%%%%%%%%%%%%%%%%%%%%%%%%%%%%%%%%%%%%%%%%%%%%%%%%%%%%%%%%%%%%

\Se{INTRODUCTION}
 
The last decade has witnessed unprecedented growth in the Internet traffic 
\cite{attinternet}. At the same time the link capacities across the Internet have also 
increased. OC-48c (2.5 Gbps) and OC-192c (10 Gbps) links are a reality now. Switches
and routers are becoming the bottleneck in fast communication. The role of 
routers have also become extremely diverse. In {\em access} networks, routers connect 
homes and small businesses to the Internet Service Providers (ISPs). Routers in
{\em enterprise} networks form the edges of the network serviced by the ISPs and 
connect thousands of computers from campuses and enterprises. The {\em backbone} 
routers at the core of the Internet connect ISPs using long distance trunks of large 
bandwidth. The requirements for these routers are different. The backbone routers 
support routing at very high rate over a few links. The enterprise routers require 
large number of ports at low cost per port, but need support for a rich set of 
value-added services. The access routers must support many ports with heterogeneous 
physical media and a variety of protocols. In addition to it, new Internet based 
applications with stringent bandwidth and delay guarantees have added to the computing 
load on a router. In this survey, we will look at the different software and hardware 
techniques employed to overcome the performance bottlenecks in a router.

Routers used to be implemented in software. A single CPU performed all the data and 
control related tasks. In a connectionless network like IP, basic data path activity
involves looking up the outgoing interface for an incoming packet. The router extracts 
the destination field from the header of the packet and uses it to perform a Longest
Prefix Match against a forwarding database of entries of the form $<network \: address/
mask,\: egress\: port >$. The address lookup involves classification based on a single 
field in the header. For providing different levels of service to individual flows, it 
is necessary to perform this classification on multiple fields in the header of a packet
to distinguish among the flows. Once the egress port is resolved the packets are routed
to the output port over a shared bus, or a non-blocking interconnect. Multiple packets
might be competing for the same output requiring a scheduling algorithm for arbitration.
In order to give service guarantees, the outgoing link must be scheduled among the flows
such that none of their guarantees are violated. This calls for a link scheduling 
discipline at each output. Each of these functions are complex operations and have to
be performed for every packet passing through the router. A software based router has the
advantage that it can be easily modified to adapt to changing specifications, but the
heavy computations associated with these operations can kill the performance. 

One way is to find better algorithms to make the operations faster. An alternative way
found its way into the router market with the advancement of IC technology. Application
Specific Integrated Circuits (ASICs) can implement many of these operations in hardware 
and achieve orders of magnitude improvement in speed. So instead of pushing the limits 
of processor speed, router designers adopted the use of specialized ASICs. The downside 
of the approach is the inherent inflexibility of the ASIC based design. Once a logic is 
burnt into silicon it is not possible to modify them. This leads to longer 
time-to-market of new designs. 

The flexibility of software based approach coupled with the speed of ASIC is the holy 
grail of router design. Network Processor (NP) is the answer to this search. Network
Processor is a programmable device that is designed and built with optimized instruction
set for networking functions. Software programmability gives flexibility of modification
and the optimized instruction set helps to reach the speed of ASIC. The term `Network
Processor' is often loosely used to refer to a collection of specialized co-processors 
for functions like address lookup, flow classification that are often integrated on the 
same chip with the programmable core.

In this paper, we survey different approaches that aim to boost router performance. On 
the software side, we will study the algorithms which have improved the runtime 
complexity of router operations. The Network Processor presents the hardware angle to 
router performance improvement. We also propose a switch scheduling algorithm which can
ensure fair sharing of the bandwidth among the flows destined to an output in an input
queued switch/router. The organization of the paper is, in Section 
\ref{Algorithms:sec} we will discuss the algorithmic solutions to the key router 
functions. Section \ref{qslip:sec} introduces our proposed algorithm for providing QoS
guarantees in an Input Queued switch. In Section \ref{architecture:sec} we point out the 
architectural choices for Network Processors and evaluate designs from some commercial 
vendors. Section \ref{benchmark:sec} looks at the need for and the means to distinguish 
among the slew of NPs currently available. We will conclude the survey in Section 
\ref{summary:sec} with our opinion. 

%###################################################################################
\Se{KEY TASKS IN PACKET PROCESSING}\label{Algorithms:sec}

Next-generation routers are expected to support a wide range of applications and 
functionalities while maintaining the high processing speed. Initially, much of the 
research had gone into speeding up the forwarding table lookup and switch fabric 
scheduling for a router. With Quality of Service in the network becoming a sought after 
requirement, research efforts were directed toward finding efficient solutions for 
packet classification based on multiple header fields, fast switch fabric scheduling 
techniques, fair scheduling of the output link and innovative memory architectures to 
prevent packet buffering and queueing from becoming a bottleneck. In this section, we 
will take a look at each of these issues and existing solutions.

\Sse{Packet Classification}\label{classification:sec}
% problem description ... multifield classification. give examples.  
% in a NP, what are the important metrics.
% talk about the different algos in relation to these metrics.

\emph {Problem Statement}: Given a database of filters or rules, packet classification
is the task of determining the filter that `best' matches the packet. 
Suppose there are {\em $K$} fields in the header that must be matched for classification. 
Then, each filter $F$ is a {\em $K$}-tuple (F[1], F[2], ..., F[k]), where each $F[i]$ is 
either a variable length prefix bit string or a range. For example, the filter 
$F$ = (130.245.*, *, TCP, 23, *), specifies a rule for traffic flow addressed
to subnet 130.245 with TCP destination port 23, which is used for 
incoming Telnet connections. If the firewall database includes this rule it will 
be able to isolate all incoming Telnet connections, and allow or discard it. 
Each filter rule in a filter database comprises an array of $K$ distinct fields, 
where $F[i]$ is a specification for the $i^{th}$ field, and an associated cost. 
The {\em best} match for a packet is decided by, (a) the order of occurrence of the 
rule in the filter database, (b) assignment of priorities to different filters, 
(c) assignment of priorities to fields to break the ties in case of multiple matches. 
For example, in firewall databases, option (a) is adopted. 

A packet can {\em match} a filter rule $F$ in three ways,
\begin{itemize} 
\item {\em Exact match} means matching with the specific value of the field in a 
filter rule.
\item {\em Prefix match} requires that the value specified in the filter is a prefix 
of the value in that particular field in the packet header. 
\item {\em Range match} is the case when the filter value is specified as a range, for 
example when specifying a set of port numbers. Any range specification can, however be
converted to a prefix-based specification. It has been shown by Feldman and Muthukrishnan
that a $W$-bit range can be represented by at most $(2W - 2)$ prefixes \cite{classify4}. 
\end{itemize}

Of the three types, Prefix Matching is the most commonly used. The two metrics
on which the algorithms in this section will be evaluated are, 
\begin{itemize}
\item {\em Space Complexity:}
The space complexity gives an estimate of the memory footprint required
to store the data structures. The bigger the data structures more difficult
it is to use faster and more expensive memory technologies to store them,
thereby increasing the memory access time.
\item {\em Time Complexity:}
The time complexity gives a bound on the maximum number of steps or cycles
that are required to perform the classification. This is directly related 
to the search speed of the classification algorithm. 
\end{itemize}

%####################################################################################
\begin{table*}
\centering
\begin{tabular}[h]{||l|c|c||}\Hl\Hl
{\bf Rule} & {\bf F1} & {\bf F2}\\\Hl\Hl
$R_1$   & 00* & 00* \\\Hl
$R_2$   & 0*  & 01* \\\Hl
$R_3$   & 1*  & 0* \\\Hl
$R_4$   & 00* & 0* \\\Hl\Hl
$R_5$   & 0*  & 1*  \\\Hl\Hl
$R_6$   & *   & 1*  \\\Hl\Hl
\end{tabular}
\caption{\sl Rule Set for Classification} 
\label{rules:tab}
\end{table*}
%####################################################################################

To enable fast processing on the data path, the most useful algorithms are the ones with the
minimum memory usage and the fastest lookup time. The most naive approach in solving the 
classification problem is to use {\em linear search} on all the rules in the filter
set. The worst-case storage and lookup time complexity for such a scenario will be 
$O(N)$. Linear increase with the number of rules presents poor scaling property. Another
hardware option is to use a Ternary Content Addressable Memory (TCAM). In this case,
the lookup time is constant, $O(1)$, but space complexity remains $O(N)$. 
TCAMs are expensive memories with limited capacities. Hence in routers where
thousands of rules are stored, it is not a viable solution.
The theoretical bounds for the classification problem is obtained from the domain 
of computational geometry. It can be framed as a point location problem in 
multi-dimensional space \cite{compgeom}. The best bounds for point location in 
{\em N} rectangular regions and $d>3$ dimensions are $O(\log N)$ time with $O(N^d)$ 
space, or $O((\log N)^{d-1})$ time with $O(N)$ space. This poly-logarithmic time 
bounds are not practical for use in high-speed routers. For example, with 1000 
rules and 5 fields, either the number of memory accesses is 10,000 per packet, or the 
space requirement is 1000 GB.

%#################################################################################### 
\myfig{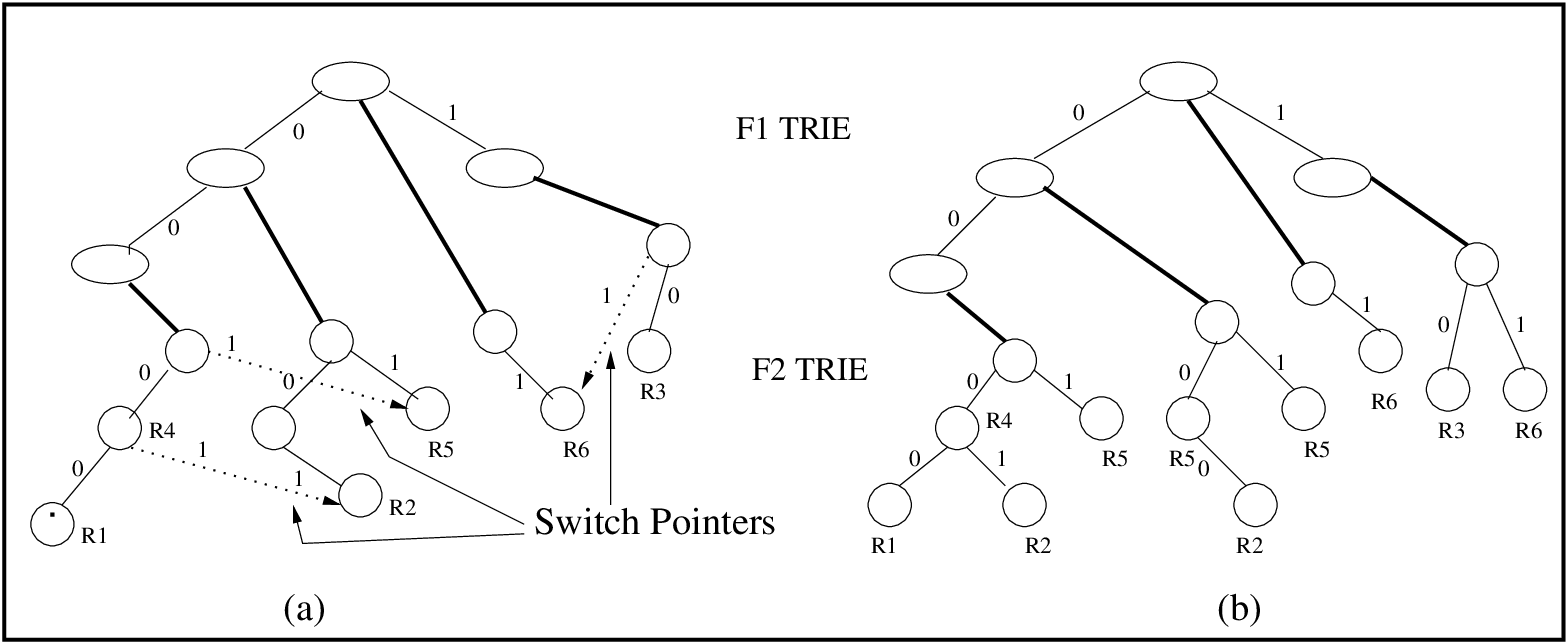}{4.0in}{\sl Examples of Trie Data Structure. Figure (a) shows a Hierarchical 
Trie data structure which represents fields F1 and F2 from the Table \ref{rules:tab}. 
Extension to Grid-of-Tries with the use of Switch Pointers is also shown. Figure (b) gives
an example of a Set Pruning Trie.}
%#################################################################################### 

As stated earlier, any range specification can be converted to prefix-based 
specification, therefore for the rest of this section, without loss of generality 
we will refer to the rule database in Table \ref{rules:tab} throughout the discussion. 
In Table \ref{rules:tab}, $R_j$ refers to a rule and $R_{j1}$ and $R_{j2}$ refers 
to the two fields for each rule. The most popular data structure for performing
fast lookup is {\em Patricia Trie} \cite{lookup1}, which is a binary branching tree with 
each branch labeled 0 or 1. Extension of a 1-dimensional trie to multiple dimensions
is done by building the trie for dimension-1 (F1) and each valid prefix in F1 points
to a trie containing the prefixes of dimension-2 (F2). This is shown in the Figure 
\ref{tries:fig}(a). The storage cost for this trie is O$(NdW)$, where $N$ is the number 
of rules, $d$ is the number of dimensions and $W$ is the maximum bit length of a prefix.
The lookup is done in a depth-first traversal which gives a lookup complexity of
O$(W^d)$. Instead of a DFS approach, breadth-first search (BFS) is proposed by
Varghese et al. \cite{classify9}, which can reduce the runtime memory requirement for the 
algorithm. However, the linear complexity for space and polynomial complexity for
lookup time is not ideal for high speed data forwarding.
A way to speed up the search procedure in Hierarchical Tries was proposed in 
\cite{classify5} by preventing backtracking with the use of {\em Switch Pointers}. Switch 
Pointers are edges labeled 0 or 1, and directs the search process to the next trie 
without backtracking. This scheme can only work in 2-dimensional classification where 
the lookup time is reduced from $O(W^2)$ to $O(W)$. Yet another way of speeding up the 
lookup can be by replicating rules to avoid backtracking. This is called a {\em Set Pruning 
Trie} and is shown in Figure \ref{tries:fig}(b). The storage complexity blows up to 
O$(N^ddW)$. In summary, use of tries leads to polynomial complexity solutions either in 
terms of space or query processing. Another solution to the multi-dimensional classification
problem was proposed by Srinivasan et al. \cite{classify7}, which is called 
{\em cross-producting}. The basic idea is to do a 1-dimensional lookup for each chosen 
field in the packet, and then compose the results to get the entry from a precomputed table 
of lowest cost filters. The table computation involves generating all possible combinations 
of filters by cross-producting individual prefixes in each dimension. The query processing 
time is reduced to O($N$ * 1-D $lookup \: time$), but the cross-product table can have a 
whooping $N^d$ entries.  

Attempts to solve the most general cases of the multi-dimensional classification 
problem lead to expensive worst-case solutions. But worst-case behavior is not a frequent 
occurrence in real world classifiers. There is considerable amount of structure and
redundancy which can be exploited to come up with intelligent heuristics as we will 
find out. We study three heuristic based solutions here.  

%#################################################################################
\begin{table*}
\begin{minipage}[t]{3.5in}
\centering
\begin{tabular}[h]{|l|c|c|}\Hl\Hl
{\bf Rule} & {\bf Specification} & {\bf Tuple} \\\Hl\Hl
R1 & (00*, 00*) & (2, 2) \\ \Hl
R2 & (0**, 01*) & (1, 2) \\ \Hl
R3 & (1**, 0**) & (1, 1) \\ \Hl
R4 & (00*, 0**) & (2, 1) \\ \Hl
R5 & (0**, 1**) & (1, 1) \\ \Hl
R6 & (***, 1**) & (0, 1) \\ \Hl
\end{tabular}
\caption{\sl Mapping from RulesSet to Tuples}
\label{tss1:tab}
\end{minipage}
\hspace{.03in}
\begin{minipage}[t]{3.5in}
\centering
\begin{tabular}[h]{|l|c|}\Hl\Hl
{\bf Tuple} & {\bf Hash Table Entries} \\ \Hl\Hl
(0, 1) & \{R6\} \\ \Hl
(1, 1) & \{R3, R5\} \\ \Hl
(1, 2) & \{R2\} \\ \Hl
(2, 1) & \{R4\} \\ \Hl
(2, 2) & \{R1\} \\ \Hl
\end{tabular}
\caption{\sl Hash Table Entries based on Table \ref{tss:tab}.}
\label{tss2:tab}
\end{minipage}
\end{table*}
%#################################################################################

A simple observation that in most real rule sets the number of {\em distinct field lengths 
used} is always small is the key to the algorithm, called {\bf Tuple Space Search} proposed 
by Srinivasan et al. \cite{tss}. The entire rule set is broken up into sets, called {\em 
tuple space} based on the length of the individual prefixes in each rule. Each 
d-dimensional rule maps to one of the d-tuples, where the $i^{th}$ field of the tuple is 
the length of the $i^{th}$ dimension of the rule. An example of generating such tuple space 
is shown in Table \ref{tss1:tab} and Table \ref{tss2:tab}. Each rule is now stored in a 
hash table corresponding to the set it maps to. The storage complexity in this case is still
$O(N)$, but the advantage comes from a faster query processing time in the average
case, where it has to do  $M$ hashed memory accesses, $M$ being the number 
of tuples and is definitely less than $N$. 

%#################################################################################
\begin{table*}
\centering
\begin{tabular}[h]{||l|c|c|c||}\Hl\Hl
{\bf Database} & {\bf Size} & {\bf Tuples} & {\bf Pruned Tuples} \\ \Hl\Hl
Fwal-1   & 278 & 41 & 11\\\Hl
Fwal-2   & 158 & 28 & 6\\\Hl
Fwal-3   & 183 & 24 & 7\\\Hl
Fwal-4   & 68 & 15 & 5\\\Hl
\end{tabular}
\caption{\sl The worst-case number of hash probes produced by the Pruned Tuple 
Search Method.} 
\label{tss:tab}
\end{table*}
%#################################################################################

The authors suggest an improvement over this approach, which they call {\em 
Tuple Pruning}. This heuristic is based on an observation that in real filter
databases there are very few prefixes of a given address. This observation is
utilized during a query by first doing individual longest prefix matching in
each dimension, and then searching only those tuples that are compatible with 
the individual matches. The authors have reported the effect of Tuple Space 
Pruning on some real firewall databases, as shown in Table \ref{tss:tab}.  

Srinivasan has suggested a different approach to tuple pruning, which is called
{\em Entry-Pruned Tuple Space Search} \cite{classify11}. In this case, with
each entry $E$ corresponding to a tuple, some information is maintained as a 
bitmap which tells which other tuples need to be searched instead of doing 
a linear search. Only those tuples that have at least one filter that does 
not contradict with a current match $E$ in any of the specified bits need to be 
looked at. The bitmap associated with each entry is $T$ bits long, where $T$ is the
number of tuple entries. The bitmap has to be precomputed by checking all
filters that does not contradict $E$. Though the naive approach for this
precomputation takes $O(N^2)$ time, the author has proposed ways to do it 
using only $O(T * N)$ time and $O(T * N)$ memory.

%#################################################################################
\begin{figure*}
\begin{minipage}[t]{3.3in}
\centerline{\psfig{figure=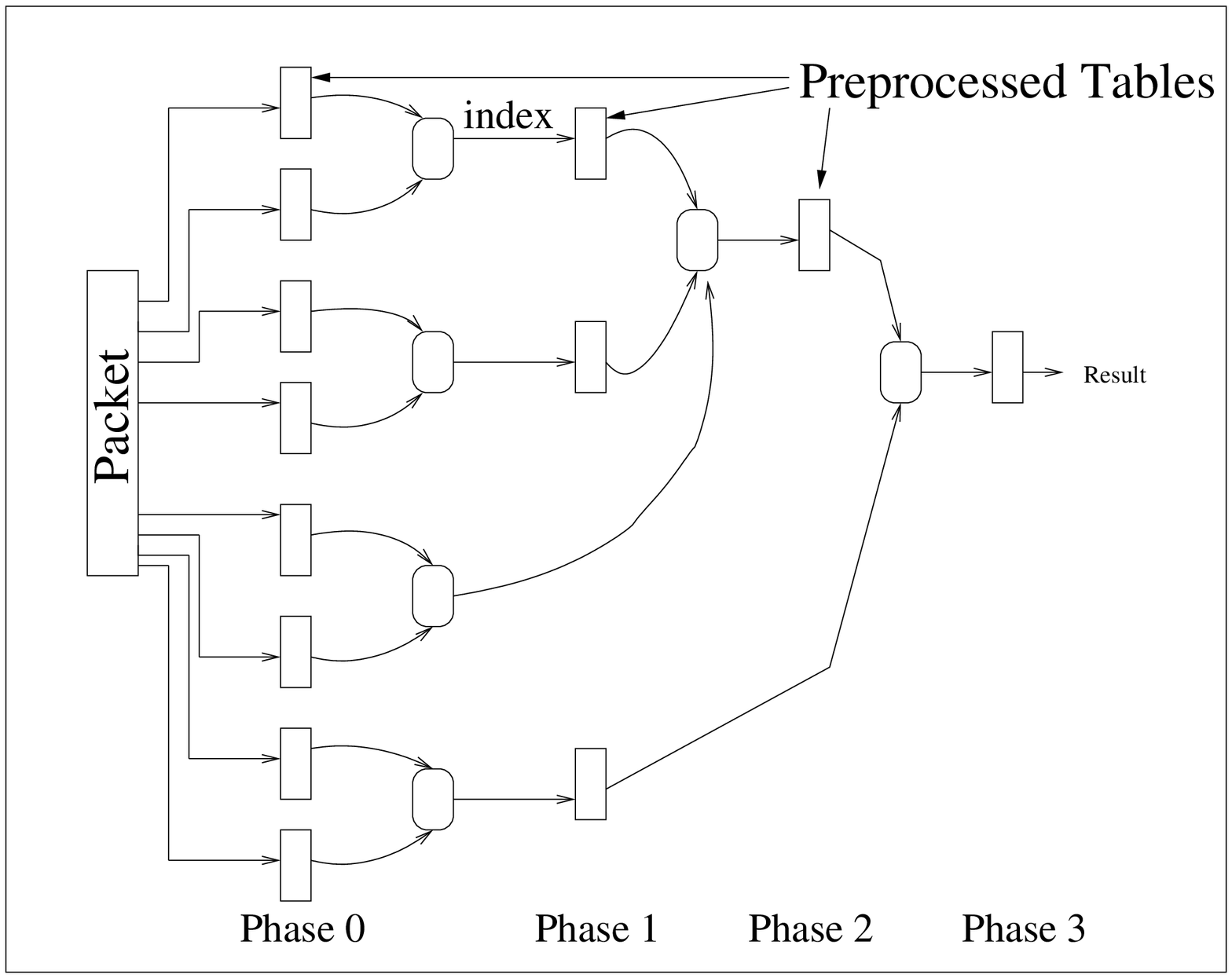,width=2.5in}}
\caption{\small{\sl Pictorial depiction of the Recursive Flow Classification Algorithm.}}
\label{rfc:fig}
\end{minipage}
\hspace{.03in}
\begin{minipage}[t]{3.3in}
\centerline{\psfig{figure=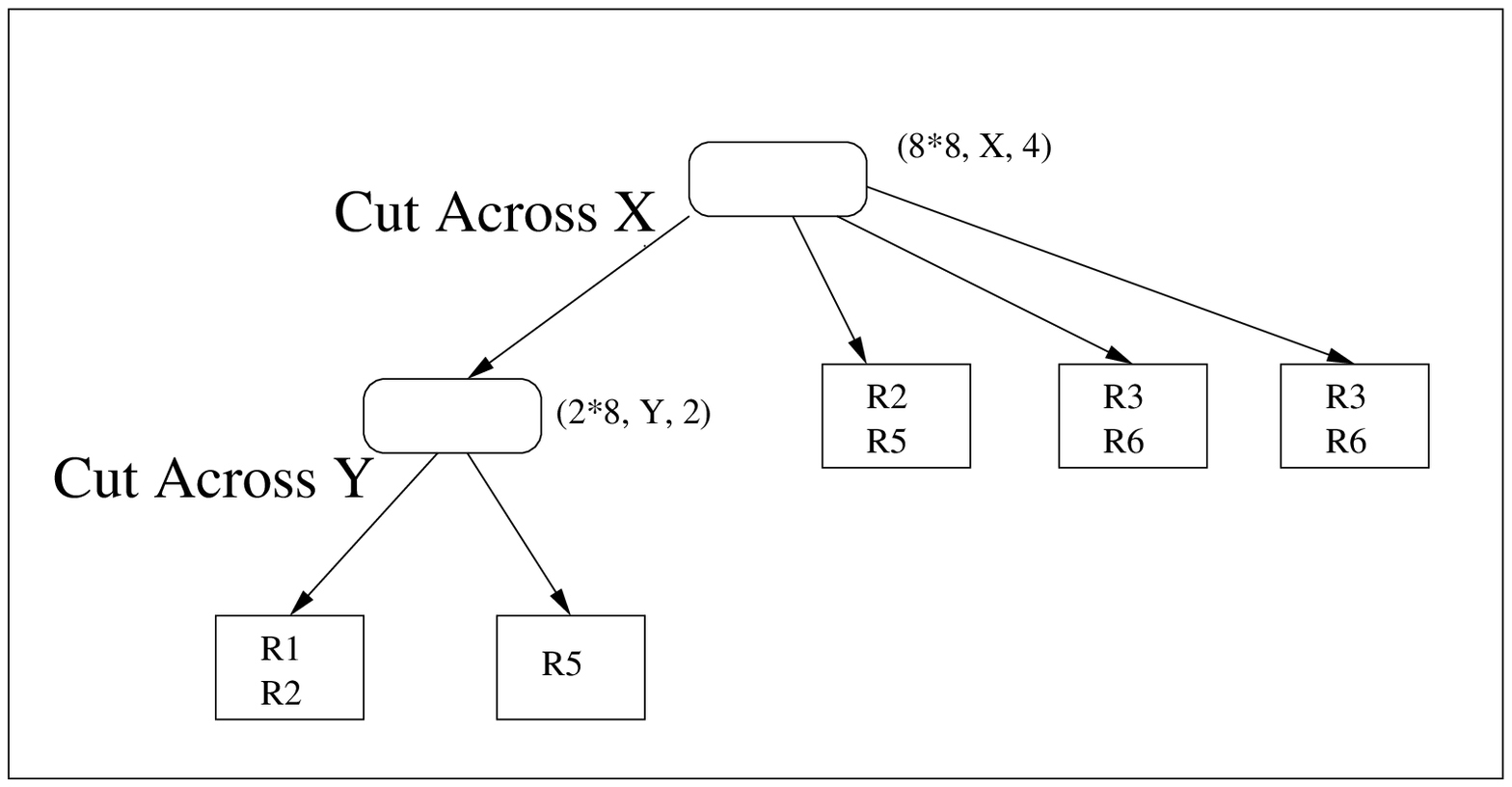,height=2.0in}}
\caption{\small{\sl A HiCut tree for the Classifiers given in the Table \ref{rules:tab}.}}
\label{hicut:fig}
\end{minipage}
\end{figure*}
%#################################################################################

The key observation for the {\bf Recursive Flow Classification} \cite{classify3} is that 
the number of overlapping regions for a dataset is considerably smaller than the worst-case 
value, which can be $O(N^d)$. Thus for a rule instead of mapping it to one of the $N^d$
possible cases one needs to search through a much reduced set. The problem therefore boils 
down to one of mapping $S$ bits in the packet header to a $T$ bit identifier, where 
$T = \log{N}$ and $T \ll S$. One impractical way for this is to precompute the value of the 
identifier corresponding to each of the $2^S$ different packet headers. This leads to a 
solution in one memory access at the expense of unreasonable amount of storage. The 
solution proposed by the authors is to do the same reduction recursively over a fixed 
number of phases. The main steps in the algorithm are,
\begin{itemize}
\item{ In the first phase d-fields of the packet header are split into 
multiple chunks of fixed number of bits that index into multiple memories 
in parallel. Each of the parallel lookups yield an output which is an
identifier. For example, let us assume we are using the 16-bit port number field 
as a chunk-size unit and the port specifications used in the rules are
\{80\}, \{20,21\}, \{$>$1023\}, \{remaining integers between 0-65536\}. Then we 
can encode this information using only 2-bits to denote the mutually exclusive
sets \{20,21\}, \{80\}, \{1024-65536\}, \{0-19, 22-79, 81-1023\}. Thus we can 
reduce from a 16-bit information to a 2-bit information based on the 
classifier rules. The key goal in each phase is to ensure that the result 
of the lookup has to be narrower than the index of the memory access.}
\item{In the subsequent phases, the index into each memory is formed by
combining the results of the lookups from the previous stage. }
\item{The last stage gives only one result which is the identifier that maps to a 
particular rule.}
\end{itemize}
A pictorial depiction of how the algorithm works is shown in Figure 
\ref{rfc:fig}. With real-life 4D classifiers of up to 1700 rules, RFC appears 
practical for 10 Gbps line rate in hardware and 2.5 Gbps in software. 

The idea behind the {\bf Hierarchical Intelligent Cut} (HiCut) \cite{classify6} algorithm, 
proposed by Gupta and McKeown is essentially partitioning the complete search space of the 
classifier rules into smaller spaces comprising fixed number of rules. The partitioning is 
based on heuristics that is determined by the structure of the classifier. Hence 
preprocessing stage that builds the search tree is the key feature in the 
solution. At the beginning of the tree building, the root node contains
all the rules. In every phase, based on the heuristics we have to choose one
dimension, {\em dim}, and the number of partitions to divide that dimension 
into, {\em np(C)}. Hence each internal node represents a subset of ranges of 
each dimension, and rules are accordingly placed in them. Finally, the 
partitioning stops when the leaf nodes have no more than a fixed number of 
rules, which can be linearly searched during a query. An example HiCut tree
based on the Table \ref{rules:tab} is shown in Figure \ref{hicut:fig}. 

The heuristics used to choose the different parameters for tree building are
explained in detail in \cite{classify6}. Intuitively, the heuristic to choose
{\em np(C)} tries to balance the trade-off between depth of the tree, giving
faster search time, at the expense of increased storage. While choosing the
dimension the heuristic can be, to uniformly distribute the rules across all
the nodes, or pick a dimension that has most distinct components. Another way
to reduce the storage is to find child nodes with same set of rules, so that
redundancy in storage can be avoided. Lastly, redundancy in the tree can be 
further reduced by clipping a rule from a leaf that has lower priority than a
rule in the same set. The algorithm was tested on 40 real-life classifiers with
up to 1700 rules. It requires less than 1 MB of storage with a worst-case query 
time of 20 memory accesses.

%#################################################################################### 
\myfig{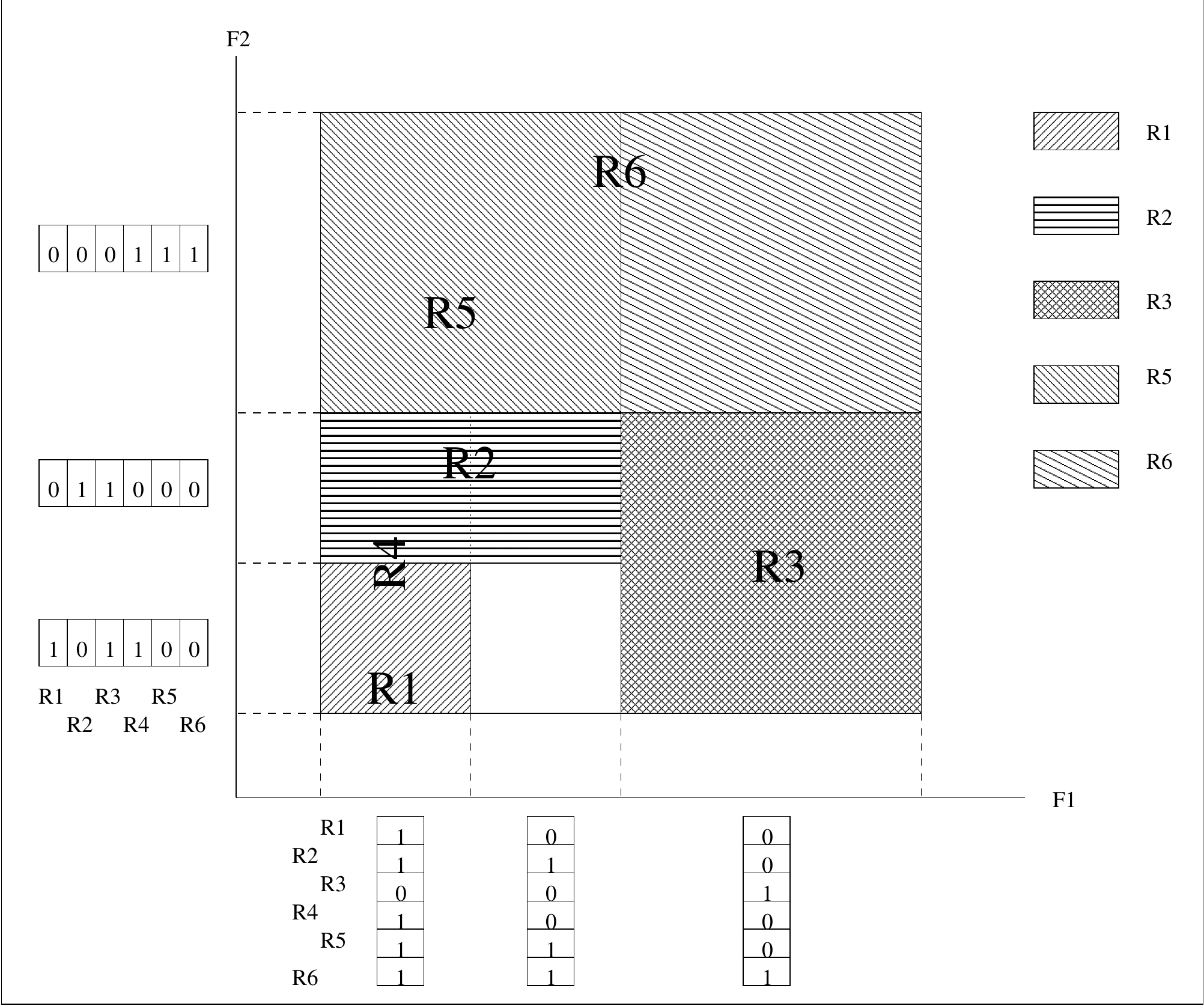}{4.0in}{\small{\sl Construction of Bitmap Tables for
Bitmap-Intersection Algorithm. In the figure the rule R4 is overlapped by the other 
rules.}}
%#################################################################################### 

In the rest of this subsection, we will look at a approach which is suited for hardware
implementation. The use of hardware can provide speedup by performing the matching in 
parallel. The Content Addressable Memories (CAMs) are suited for this purpose because it 
can do an exact key match for all the rules in parallel. But the drawback is that it must 
do an exact match and the rules has to be ordered according to priority. 
Inability to store a mask bit in Binary Cam can be overcome by using a
Ternary CAM (TCAM). But CAMs or TCAMs are limited in capacity. Hence it restricts
the number of the filter rules that can be stored. Also, larger the memory, more
power it dissipates. At present, there are companies producing 2 Mb TCAMs 
capable of single and multifield classification in as little as 10 ns. 

Lakshman and Stiliadis proposed an algorithm that proceeds by decomposing the search in 
each dimension followed by a linear complexity combining step \cite{classify1}, which is 
called {\bf Bitmap Intersection}. Although this is slower than the best 
known poly-logarithmic solutions, the authors contend that the simplicity of 
the solution makes it ideal for parallelizing and for implementing in hardware. The
algorithm first has to preprocess the filter rules to build a set of bitmaps 
that directs the search. In the preprocessing, all the intervals belonging to
a dimension, say $j$, is projected on to the $j^{th}$ axis, leading to a maximum
of $2n+1$ non-overlapping intervals. Now, for each interval, a bitmap of $N$ 
bits is created denoting if the rule belongs to that interval. Figure \ref{lakshman:fig} 
illustrates the bitmap for Table \ref{rules:tab}.

The search procedure proceeds by first selecting the packet header field for 
each dimension, and then using binary search to locate the interval it belongs to.
Next, we find the conjunction of the corresponding bit vectors in the bit arrays 
associated with each dimension and choose the highest priority entry in the resultant 
bit vector to set the solution. It is assumed that the rules are in decreasing priority. 
The overall approach being simple the only hardware elements that are required
for binary search operation is an integer comparator, and the only operation
for the intersection is parallel AND operation. The complexity of the algorithm
for search time is O$(dt_{RL} + dN/w)$ and O$(dN^2)$ in storage, where $t_{RL}$
is the time for range lookup in one dimension and $w$ is the memory word length.
The experiments with this solution provides that the scheme can support up to
512 rules with a 33 MHz FPGA and five 1 Mbit SRAM, classifying 1 Mpacket/sec.

An optimization for this algorithm is proposed by Varghese et al., which is
called {\em \bf Aggregate Bit Vector} \cite{classify12}. The main idea is that 
for each basic interval in one dimension number of rules may be few. The $N$ bit
vector formed is therefore sparse, having 0 in most bit positions. Since,
the query time of the bit-vector algorithm depends on the memory word length, so
reducing the amount of reads will make the searches faster. The idea is to compress
each $N$ bit-vector by replacing every $m$ consecutive bits with a 0 if all the $m$ 
bits are 0s. For example, a bit vector {\tt 000011110000} can be replaced by
the Aggregate Bit Vector {\tt 010}, assuming granularity of compression is 4.

%####################################################################################
\begin{table*}
\centering
\begin{tabular}[h]{||l|c|c||}\Hl\Hl
{\bf Algorithm} & {\bf Worst-case Execution} & {\bf Worst-case Storage} \\
 & {\bf Time Complexity} & {\bf Complexity} \\ \Hl\Hl
Linear Search   & N & N \\\Hl
Ternary CAM   & 1  & N \\\Hl
Hierarchical Tries  & $W^d$  & NdW \\\Hl
Set-pruning Tries   & dW & $N^d$ \\\Hl\Hl
Cross-producting   & dW  & $N^d$  \\\Hl\Hl
RFC   & d   & $N^d$  \\\Hl\Hl
HiCuts   & d  & $N^d$  \\\Hl\Hl
Tuple Space Search   & N  & N  \\\Hl\Hl
Bitmap intersection   & dW + N/memwidth  & d$N^2$  \\\Hl\Hl
\end{tabular}
\caption{\sl A comparison of different multi-dimensional classification algorithms.} 
\label{classifier:tab}
\end{table*}
%####################################################################################

The summary of the discussion is shown in Table \ref{classifier:tab}. From Table 
\ref{classifier:tab} it can be concluded that Ternary CAMs are the best solution for 
classification. If the forwarding tables in a router are reasonably small, TCAMs can be 
used for speeding up the router. But the linear storage complexity is its main downside. 
The best worst-case bound that is achievable for storage is $O(N)$ for the other approaches 
too. However, with heuristic approaches like Tuple Space Search with the same storage 
complexity we can get a better average case time complexity than linear search.

%   ii.	Memory Architectures for fast memory access to scale to 
%	high speed.
\Sse{Buffer Management Schemes}\label{buffer-mgmt:sec}
% What is the need for buffering ? 
% What was the conventional and the easiest method ? 
% Central Shared Buffer Memory : merits and demerits, similar to output queueing
% Alternative: Input Queueing 

Buffering of packets is a key operation in a packet switch. A fast switch/router design
needs to take into account the buffer management schemes so that the memory bandwidth 
does not become a bottleneck in high speed processing. The ways to overcome memory
bottleneck using existing DRAM and SRAM technologies involve, (i) coming up with
intelligent buffering schemes that reduce the aggregate memory bandwidth requirement,
and/or (ii) designing novel memory architectures that can reduce memory access time.
In this subsection, we will first take a look at the various buffering schemes, pointing
out the trade-offs for each of them, and then discuss two novel memory architectures that 
enable fast packet buffers.

The simplest buffer design is to connect the line cards to a shared backplane and write
all arriving packets to a pool of shared buffer memory where they wait their turn to be 
sent out on the outgoing link. A shared memory architecture is attractive because it can
achieve 100\% throughput, minimize average queueing delay and makes it easier to 
guarantee quality of service. But the shared backplane or the shared buffer memory 
must have sufficient bandwidth to accept packets from and write packets to all the line cards
in a single time slot. In other words, the shared bus and the shared memory for a router 
with $N$ line cards each connected to a line of rate $R$, must have a bandwidth of $2NR$. 
For a 10 Gbps (OC-192c) link and a 16 port switch, the bandwidth required is 320 Gbps.
Existing memory technologies cannot support this rate. This design in essence is 
equivalent to {\em Output Queueing} (OQ), where when a packet arrives it is immediately 
placed in a queue that is dedicated to its outgoing link. One choice is to have 
dedicated memory on each line card for holding packets that will go out through this 
outgoing link. The individual memory bandwidth requirement in one time slot is $(N + 1)R$ 
to allow for $N$ writes from the $N$ inputs and 1 read for sending out a packet.

An alternative is that the buffers on each line card hold every packet 
on {\em arrival} at the input port. Every time slot a non-blocking switch fabric 
(crossbar switch) must direct the packets to its output port. This buffering scheme is
called {\em Input Queueing} (IQ). This requires the bandwidth of each line card memory to be 
$2R$ and that of the switch fabric to be $NR$. Input Queueing can suffer severely from 
head-of-line (HOL) blocking. If there are packets at the head of the queue destined 
to an output which is busy, then packets later in the queue cannot be sent out even if 
its outgoing link is free. This can severely affect the throughput. If each input maintains
a single FIFO, then HOL blocking can limit the throughput to just 58.6\% \cite{islip}. 
One way to improve the throughput is to increase the speed of the switch fabric by a 
factor of $S$, where $S$ denotes the number of packets that can be transferred in one 
time slot\footnote{One time slot is the time between packet arrivals at input ports.} 
from the input to the output. If $S = N$, it is equivalent to Output Queueing. However, even 
with $S = 1$, it is possible to achieve 100\% throughput if we use {\em Virtual Output 
Queueing}, in which each input maintains a separate queue for every output. 

Input Queueing can make switches run faster, but Output Queueing is the design of choice 
for implementing packet scheduling schemes\footnote{ We will look at them in Section 
\ref{psched:sec}} with guaranteed QoS. The challenge therefore is to come up with a 
buffer management mechanism such that we can get identical behavior as OQ-ing without 
incurring the high memory bandwidth requirements associated with it. The two values,
1 and $N$ of the parameter `speedup' ($S$) presented two ends of the spectrum. If we can
buffer packets both at the inputs as well as at the outputs after switching, then different
values of $S$, $1 \leq S \leq N$, should give different throughputs. This scheme is called
Combined Input-Output Queueing (CIOQ). In CIOQ the challenge is to determine
the smallest value of $S$ and an appropriate cell\footnote{ Packets are usually 
segmented into fixed-sized cells before switching and reassembled before departure}
scheduling algorithm. It was shown by Chuang et al. \cite{cioq} that a $N$ x $N$ CIOQ switch 
with a speedup of 2 can exactly emulate a $N$ x $N$ OQ switch with either FIFO scheduling or 
any other scheduling, like WFQ or strict priority. The key to the result is the scheduling
algorithm, which decides the order in which the cells at the input are transferred 
across the switch fabric to the output in such a way that the cells are sent out on the 
output link at the same time as in an OQ switch. This in turn is dictated by the order in
which the cells will be inserted into the input queue. So in \cite{cioq}, the authors have
described an insertion policy, called Critical Cells First (CCF). The intuition is to defer
the transfer of an arriving cell across the switch as long as possible by inserting it as
far from the head of its input queue as possible. For this it calculates the priority of 
the cell at the output port, $OP(c)$. Assuming the $OP(c) = X$, the cell is inserted at 
the $(X+1)^{th}$ position in the input list. The input priority, $IP(c)$ is thus set to 
$(X+1)$. Now, slackness, $L(c)$ is defined as $OP(c) - IP(c)$. One time slot is
broken up into 4 phases: arrival of cells at the input, scheduling of cells from input to 
the output, departure of cells from the outputs, and the second schedule. This requires
a speedup of 2. The slackness can never decrease as a net effect of the 4 phases, therefore,
slackness cannot decrease from one time slot to another. This means that when a cell is 
supposed to leave, i.e. the $OP(c)$ has become zero, then $IP(c)$ must also be zero. The
cell is thus at the head of the input and the output priority list. In this case, the stable 
matching algorithm is guaranteed to transfer it to its output during the time slot, and 
therefore the cell departs on time.
    
%#####################################################################################
\myfig{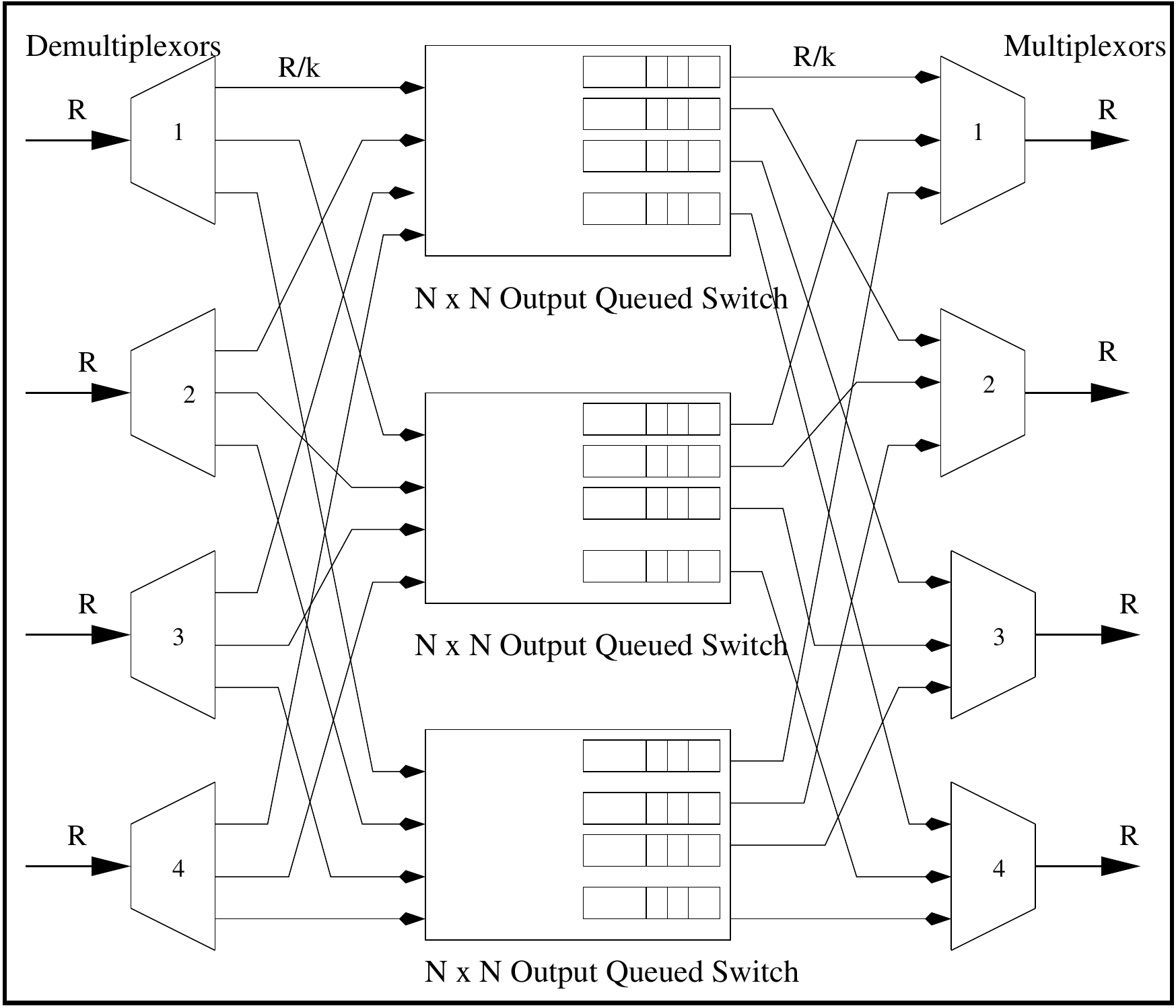}{3.0in}{\sl Architecture of the Parallel Packet Switch with a single stage of 
buffering at the center stage. The buffer memories can run at speeds of $R/k$, where $R$ 
is the line rate and $k$ is the number of center stage OQ switches.}
%#####################################################################################

A separate class of buffering schemes, where only one stage of buffering is used, unlike
CIOQ (IQ and OQ can also be considered in this category) have been modeled and analyzed 
in a recent work by Iyer et al. \cite{sbrouter}. Single stage buffering covers a wide 
variety of schemes starting from the use of Parallel Shared Memory to Distributed Shared 
Memory for buffering. A novel example of the SB architecture is the Parallel Packet Switch 
(PPS) design \cite{pps}, shown in Figure \ref{pps:fig}. The architecture is based on 
output-queued switches and resembles a Clos Network with buffering present only at the 
center stage. The main idea is to use memories running at a slower pace than the line rate 
by doing load balancing or inverse multiplexing across the slower switches. The internal
links now run at rate (R/k), where R is the external line rate and k is the number
of internal stages. Again, the goal was to mimic the FCFS-OQ switch and an OQ switch
that can support different scheduling disciplines. It was shown that a speedup of 2
is needed to achieve the first goal, and a speedup of 3 is required for the second.
The analysis is based on understanding the constraints while choosing the intermediate
layer to buffer the packet. The layer chosen should be such that no other packet is 
being written to it and no other packet is being read from it at the same time when the 
packet will be written to it or when it is time for the packet to be read out. This 
constraint set technique of analysis has been dealt in detail in \cite{sbrouter}.   

%####################################################################################
\begin{table*}
\centering
\begin{tabular}[h]{||l|c|c|c|c||}\Hl\Hl
{\bf Buffering} & {\bf \# of memories } & {\bf BW of} & {\bf Total BW} & {\bf Crossbar BW} \\
{\bf Schemes} & {\bf memories} & {\bf each memory} & & \\ \Hl\Hl
Shared Memory & 1 & 2NR & 2NR & \\\Hl\Hl
Output Queued & N & (N+1)R & N(N+1)R &  \\\Hl\Hl
Input Queued & N & 2R & 2NR & NR \\\Hl\Hl
CIOQ (speedup of 2) & 2N & 3R & 6NR & 6NR \\\Hl\Hl
PPS (FCFS-OQ) & kN & 2R(N+1)/k & 2N(N+1)R &  \\\Hl\Hl
PPS (WFQ-OQ) & kN & 3R(N+1)/k & 3N(N+1)R &  \\\Hl\Hl
\end{tabular}
\caption{\sl A Comparison of different buffering schemes.}
\label{buffering:tab}
\end{table*}
%####################################################################################

Table \ref{buffering:tab} summarizes the requirements of the different buffering schemes
discussed so far. Although output queueing and shared memory architecture presents the
theoretically best solution, it is infeasible in terms of memory bandwidth requirement.
CIOQ gives the best balance between bandwidth requirement and reaching the goals of 
guaranteed service.  

%#####################################################################################
\begin{figure*}
\begin{minipage}[t]{3.5in}
\centerline{\psfig{figure=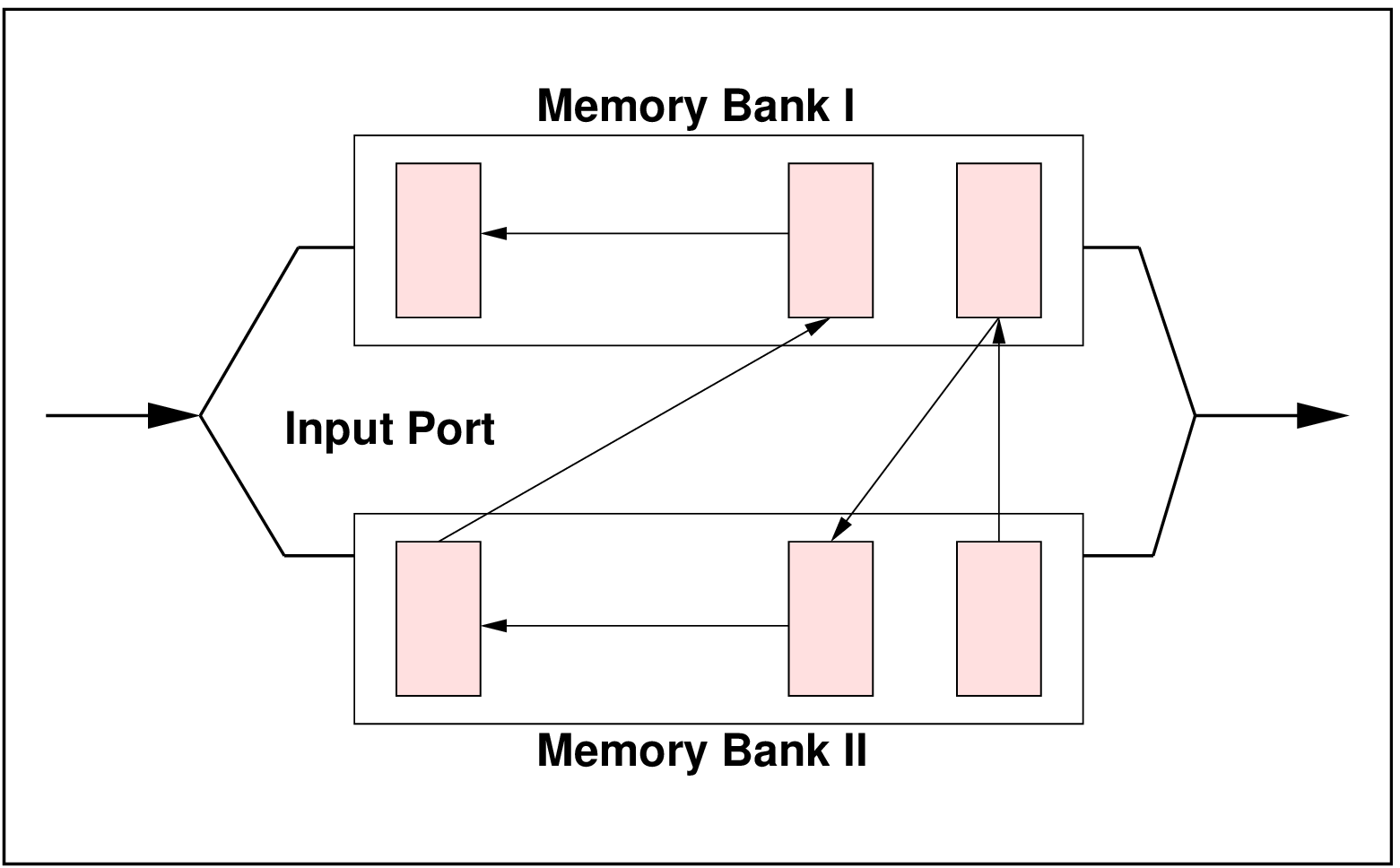,width=3.0in}}
\caption{\small{\sl Schematic diagram showing how ping-pong buffering
technique works on a port of an input-queued switch. The arrows denote the
order in which the packets are written to each memory bank on arrival. If one
bank is busy with a read operation, then an arriving packet is written on to
the other memory bank.}}
\label{pingpong:fig}
\end{minipage}
\hspace{.03in}
\begin{minipage}[t]{3.5in}
\centerline{\psfig{figure=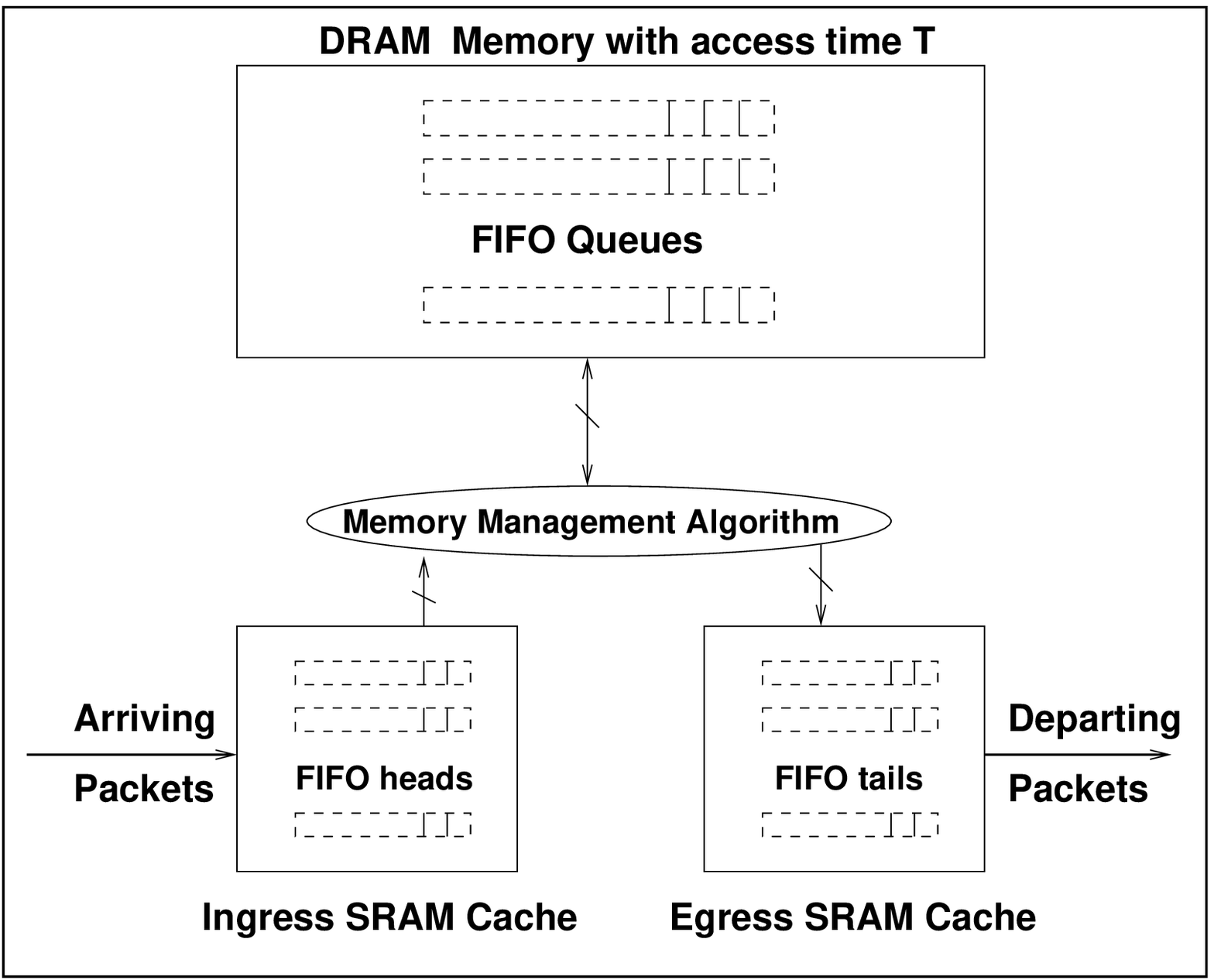,width=3.0in}}
\caption{\small{\sl Memory hierarchy used in the design of ECQF-MMA design
for fast packet buffers. In this design, the ingress SRAM is used to hold the 
recently arrived packets and the egress SRAM holds the packets that might be 
requested in near future. The MMA algorithm decides the writing/reading into 
the larger DRAM from the ingress/egress SRAMs.}}
\label{ecqf:fig}
\end{minipage}
\end{figure*}
%#####################################################################################

Unlike PPS, which makes use of memories running slower than the line rate, the following
efforts were directed at designing faster memory architectures. Joo and McKeown proposes
the idea of {\em ping-pong buffering} \cite{ping-pong}, which is similar to
memory interleaving. It uses two conventional single-ported memories and
allow only one operation to each physical memory in a time slot. This leads
to an effective doubling of the memory bandwidth. The architecture for 
ping-pong buffering with cells queued in it is shown in Figure \ref{pingpong:fig}. 
The scheme might lead to a state when even though some memory is available it can 
cause an overflow. It happens on a packet arrival when one bank is full, and the other 
non-empty bank is being read. It was shown through simulations that in the presence of
bursty traffic such conditions are frequent and  authors claim that an extra 5\% of 
memory can prevent this from happening.

Another innovative scheme for speeding up packet buffers, called the 
{\em Earliest Critical Queue First - Memory Management Architecture} (ECQF-MMA) 
\cite{ecqf} is proposed by Iyer et al. They come up with an intelligent scheme 
of building a buffer memory comprising large, slow, and low-cost DRAMS along 
with small, fast but more expensive SRAMS. The architecture is shown in 
Figure \ref{ecqf:fig}. Here the SRAMs act like a cache and holds the tail 
(ingress) and the head (egress) of each FIFO queue present in the switch. When
$b$ cells have accumulated in the ingress SRAM it is written to the DRAM, 
and when a read request is issued $b$ cells are read at a time from the 
DRAM. It is shown that for this scheme to work the minimum SRAM 
size required is $Q(b-1)$, where $Q$ is the total number of FIFO queues. The 
authors also prove that any sequence of requests from a scheduler, e.g. a 
switch scheduler can be serviced with a bounded delay of $(Q(b-1) + 1)$ time 
slots. This is done by using a {\em lookahead list} of service requests and 
issuing reads to the DRAM for queues that will be drained from the SRAM 
earliest.

%   iii.  Switch scheduling.

% CIOQ switch scheduling ... talk about how to place the packet in the input queue (CCF)

\Sse{Switch Fabric Scheduling}\label{switch-sched:sec}
In the last subsection, we saw that buffering incoming packets at the input is a 
popular design decision because of the reduced memory bandwidth. These packets must be 
routed to their respective output ports at the `correct' time. The inputs and the outputs 
are usually connected by a non-blocking switch fabric like a crossbar interconnect. 
Every time slot the crossbar must be configured to match a non-empty input to an 
output. The goal of a scheduling algorithm is to maximize the switch throughput by 
coming up with the best possible match. For buffering schemes which need to provide 
guaranteed service controlling the delay of individual packets is also important. We 
will see how, in case of CIOQ, the choice of inserting a packet into the input queue 
can satisfy the second goal and result in different scheduling algorithms.  

The important metrics for judging a switch scheduling algorithm are, 
\begin{description}
\item{\em Efficiency}, which is determined by the throughput of the switch 
using this algorithm. The throughput of a switch is defined as the number of 
packets that the switch can transfer in one time slot.

\item{\em Fairness}, which ensures that the algorithm services packets from every
input queue in a fair manner without starving any queue.

\item{\em Implementational Simplicity} requires that the algorithm be suitable 
for hardware implementation. This is a useful feature for high speed switch 
design where millions of packets need to be scheduled through the switch core 
every second. 
\end{description} 

%###################################################################################
\myfig{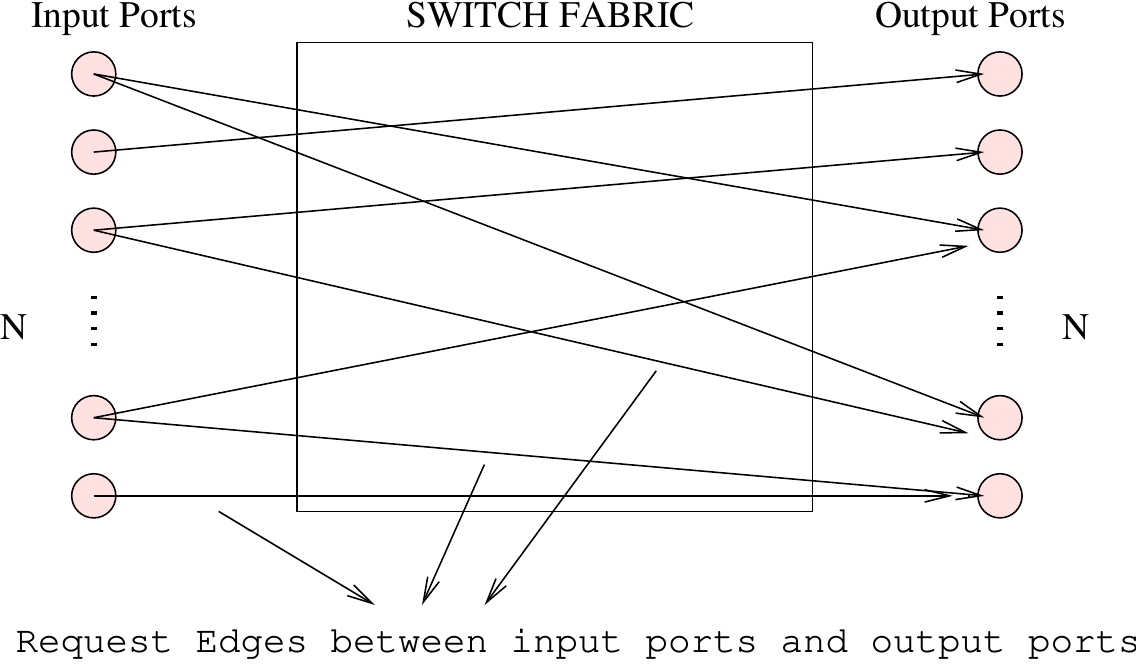}{3.0in}{\small \sl The diagram shows the equivalence of the switch fabric
scheduling with the bipartite graph matching problem.}
%###################################################################################

In a non-blocking crossbar interconnect $N$ inputs are forwarding cells or fixed-sized 
packets to the $N$ outputs, where more than one packet can be destined for the same 
output. The problem of scheduling can be modeled as a matching problem in a bipartite 
graph, as shown in Figure \ref{bipartite:fig}. {\em Maximum size matching} or 
{\em maximum weight matching}, if weights can be assigned to the edges, gives the 
optimal configuration for a time slot. The weights can be some sort of a state 
indicator for the input, for example, the length of the queue at the input, or the 
age of the oldest cell held at the input. The complexity for maximum 
weight matching (MWM) is $O(N^3logN)$ \cite{bi-match1}, and for maximum size matching 
(MSM) it is $O(N^{5/2})$ \cite{bi-match2}, where $N$ is the number of inputs and outputs. 
The execution time is too high for use in high speed packet switching. The main research
focus has been to come up with heuristics that can solve MSM or MWM in a fast and 
easily implementable manner. 

A comprehensive survey of a number of switch scheduling algorithms is presented
in \cite{sw-survey1}. For completeness, we will highlight the major solutions, and 
discuss developments of an emerging class of scheduling schemes based on {\em 
randomized algorithms}. The three main steps in switch scheduling as pointed out 
in \cite{sw-survey1}, are (i) the computation of the weight matrix (in cases where 
the algorithm tries to approximate the MWM solution, otherwise for MSM the weight of 
all the edges are assumed equal), (ii) selection of the heuristic scheme, (iii) the 
resolution of contention among inputs/outputs. 

A set of algorithms that closely matches throughput of the MSM approach are,
\begin{itemize}
\item{\em Parallel Iterative Matching} (PIM) \cite{pim}, developed by DEC 
Systems Research Center, uses random selection among inputs and outputs to 
ensure that none of the inputs are starved for a share of the switch fabric, which is one 
of the main pitfalls in the MSM solution for scheduling. The algorithm works in 3 basic 
steps,
\begin{description}
\item[STEP 1.] {\it Request}, where each unmatched input sends a {\em request} to 
every output for which it has a cell to send.
\item[STEP 2.] {\it Grant}, where an unmatched output \underline{randomly} selects 
one of the inputs who has sent a request to it, and replies with a {\em grant}.
\item[STEP 3.] {\it Accept}, where an input that has received grants from 
more than one output will \underline{randomly} pick an output among the ones which 
has sent back grants. 
\end{description}
In \cite{pim}, the authors have shown that the algorithm converges to a maximal
match in $O(logN)$ iterations. But in fast switches, number of iterations often 
need to be restricted to just 1. For 1 iteration PIM gives a throughput of 
63\%. 

\item{\em 2 Dimensional Round Robin} (2DRR) is different from PIM in the 
resolution of the conflict among contending inputs and outputs. Each output in 
2DRR maintains a round-robin arbiter for the inputs and vice-versa for choosing the output 
at each input. 
When a request arrives at an output, it `grants' the input that appears next in a fixed, 
round-robin schedule starting from the highest priority element, and the pointer in
the round-robin table is advanced to the next element. The input takes a similar 
approach in choosing which output to send out the packet to by picking the highest
priority element from the round-robin list of the outputs. This algorithm fails to
give good throughput at high loads because of the synchronization of the grant pointers.
When inputs are saturated with packets for all the outputs, the grant pointers advance 
in lock-step, leading to a maximum throughput of 50\%.   
 
\item{\em iSLIP} algorithm \cite{islip}, proposed by McKeown et al. improves
on Round Robin Scheduling to achieve 100\% throughput under uniform saturated load. 
It breaks the synchronization by advancing the grant pointer only if the grant is 
accepted by the input. 
\end{itemize} 
The algorithms discussed so far have iterative solutions. In actual implementation,
PIM executes 4 iterations in each time slot. With more iterations there is an
improvement in throughput for all these algorithms, as shown in a quantitative
survey in \cite{sw-survey2}.

Another class of algorithms for scheduling try to get approximate solutions to the 
MWM problem. The MWM solution provides better fairness among the inputs. The algorithms 
in this category differ in the computation of the weights associated with each edge in 
the input graph. The weights can be decided by different metrics at the input or output 
side. In the {\em Oldest Cell First} (OCF) algorithm \cite{ocf} the waiting time of 
the Head-of-Line cells is used as the weights in the weight matrix. The algorithm 
tries to find the Maximum Weight Match based on this weight matrix. In the 
{\em Longest Queue First} (LQF) algorithm \cite{lqf} the weight of an edge in the 
weight matrix represents the number of cells held in each queue at the input. It uses 
this Weight Matrix to solve the MWM problem. This is also an iterative solution. 
Though this is a simple and stable solution, iLQF can lead to starvation in some cases. 
Moreover, it was pointed out that implementing LQF in hardware is too complex. 
The {\em Longest Port First} (LPF) was suggested by McKeown et al. to overcome
the complexity problems of LQF. The key idea of the LPF algorithm is to service
the queues with the highest port occupancy. At every time slot, say at the $n^{th}$ 
slot, port occupancy is calculated as 
$R_i(n) + C_j(n)$, 
where $R_i(n)$ is the total number of cells currently buffered at input $i$ and 
$C_j(n)$ is the total number of cells at all inputs waiting to be forwarded to output $j$. 
Although weights are used, it is proved that LPF can be reduced to a MSM 
problem and an iterative solution suitable for hardware implementation is easy
to formulate.  

Using {\em randomized algorithms} to come up with the best match is another approach.
The basis for this method stems from the observation that there is some
sort of temporal correlation in the sequence of matchings. The matching at one
time slot is generated as a random variation of the matching in the previous
time slot because it can be safely assumed that under moderate or high loads
the weighing factors will not change from one match to the next. The idea was
shown to work by Tassiulas \cite{random1}. The approach can be summarized as,
at time $t+1$ choose a matching $R$ at random from a set of
$N!$ possible matchings, and compare it with the previous matching. If it has a 
weight greater than the previous matching use it as the current match. This
approach achieves a maximum throughput of 100\%, but can lead to high delays.
Prabhakar et al. improved on Tassiulas's algorithm by noting that a small number
of edges, say $m$ $(m < N)$ contributes to the bulk of the weight of a matching. 
So keep these $m$ edges and pick the remaining $(N - m)$ edges randomly. As explained 
in their algorithm, called LAURA \cite{random2} they use the matching at time 
$t-1$ to select a set of edges that contributes to a {\em proportion of the
total weight} and then selects the remaining input/output matches randomly. 
These {\em randomized algorithms} are appealing for their simplicity because 
they can achieve a complexity linear with the number of input/output ports.   

The algorithms discussed so far are focused at increasing the switch throughput, but 
do not talk about giving a QoS guarantee to the flows. With buffering at the input, 
scheduling algorithms need to take care of that as well. One way is to distinguish 
among flows while choosing a match. All connections from an input can be bundled into 
one group and provided a fixed share on the output. This is done in the Weighted 
Parallel Iterative Matching (WPIM) \cite{wpim} algorithm by Varma and Stiliadis. It 
augments the PIM algorithm with an extra step of {\em masking} before {\em grant} 
replies are sent back. In this scheme, each input-output connection requests a 
{\em credit} which is the fraction of the bandwidth desired by this input on the 
output link. Before giving a grant, those inputs which have received their share are 
ruled out from selection consideration. This gives the guarantee to every connection 
from a specific input that they will receive their share of the bandwidth. 

\mycomment{
Another possibility is to maintain the input queue in a way such that the packets are
sent out to their outputs prior to or at the time they are due for transmission. The 
Critical Cell First (CCF) algorithm used for insertion into the input queue for CIOQ switch 
with a speedup of 2 ensures that any OQ switch can be exactly mimicked. For any arriving 
packet it calculates when it is supposed to leave the switch, i.e. it must be at the head 
of the output queue at that time. When a packet arrives it is pushed into the input queue 
in such a way that it is guaranteed to be scheduled for transfer to the output before its 
departure time expires. Using a speedup of 2 ensures that this constraint always holds.   
}

%   iv. Flow Management ie. algorithms for providing QOS, COS, 
%	traffic shaping.

\mycomment{
1. output link scheduling problem ...
2. fluid fair queueing and Packetized version (WFQ) ... 
	what are the main operations and bounds.
3. Some more examples, but they cannot break the bounds ... 
4. Delay and Rate are associated .... how to break the association. 
	Rate proportional Servers.
5. Round-Robin scheduling ... DRR, SRR, Discretized FQ.
6. Distributed FQ, PILO ... meant to provide QoS on CIOQ switches.
}

\Sse{Packet Scheduling over Output Link}\label{psched:sec}
In best-effort service packets are dispatched in a First-Come-First-Serve (FCFS)
order. The integrated services, which support voice, video and other real-time data traffic,
require QoS guarantees. The shared resource, in this case the outgoing link, must be 
allocated to packets from different flows according to each flows' reservations or service 
level agreements (SLAs). In this subsection, we will discuss the class of algorithms that
have addressed the problem of scheduling packets over the outgoing link. Most
of these scheduling algorithms assume an output queued switch model. Later in this 
section, we will also look at two solutions that address the packet scheduling problem 
in the context of CIOQ switches. 

The link scheduling algorithm should be able to give guarantees on parameters like 
average throughput, end-to-end delay, delay jitter. More parameters an algorithm can
handle, more flexible is the algorithm. Another important criterion is to guarantee
that rogue flows do not disturb the well-behaved flows, or when there is leftover
bandwidth after meeting the guarantees, it is fairly distributed among the remaining 
backlogged flows. All scheduling algorithms must have a bounded fairness compared 
to the ideal case. We are dealing with millions of flows when scheduling. Hence the
algorithms should scale to tackle such large number of flows at the required high speed.
Finally, as these algorithms will run on the high speed data paths, 
implementation simplicity is a desired feature. We will see that there are two main 
classes of link scheduling algorithms, ones that are based on the conceptual model of
Fluid Fair Queueing (FFQ), and the other based on Round Robin Scheduling.  

{\em Virtual Clock} algorithm similar to Time-Division Multiplexing (TDM) was proposed 
by Lixia Zhang \cite{vclock}. The Virtual Clock algorithm takes the approach of 
timestamping each arriving packet with a virtual transmission time based on the flow's 
reservation. Packets are transmitted in the increasing order of their virtual 
transmission times. The Virtual Clock algorithm ensures that the transmission maintains 
a strict delay bound, but it can perform badly in ensuring fairness among the backlogged
queues. Lack of a system-wide parameter to keep track of the total service each flow 
would have received can lead a backlogged session in the Virtual Clock discipline to be 
starved for an arbitrary period of time. Using the excess bandwidth during periods 
when other flows had no packet to send leads to this state.

% what is Fluid Fair Queueing (FFQ) ?
The conceptual model for packet scheduling is based on a Fluid Fair Queueing Server
\cite{fq90}. Each of the $N$ flows are characterized by $N$ positive real numbers, $\phi_1, 
\phi_2,\ldots, \phi_N$, which denote the respective share of the total channel 
capacity $C$. During any time interval $(\tau_1, \tau_2]$, if there are $N$ backlogged flows
each of them are serviced simultaneously in proportion to the weights. The service
received by a flow $n$ in time interval $(\tau_1, \tau_2]$ is given by,
\begin{equation}
r_n(t) = \frac{ \phi_n }{\sum_{j\in B(\tau_2)} \phi_j }.C(t), t \in (t_1, t_2]  
\label{ffq:eqn}
\end{equation}
where $B(\tau_2)$ denotes the set of backlogged queues and $C(t)$ is
the link-speed which can be variable. In a real switch, the unit of transmission is
usually chosen as one packet. The packetized version of the FFQ server aims to 
minimize the difference in service offered to each flow compared to FFQ. There are two
main operations for any scheduler trying to emulate FFQ. First, it has to keep track
of the total service a flow has received in the FFQ. This is maintained by a system 
wide variable, called the {\em system virtual time} ( V(t) ). Second, there must be some
index to order the packets on the outgoing link. This is done by ordering the packets
based on their departure times. So we have two more variables, {\em virtual start time},
$S_i(t)$ and the {\em virtual finish time}, $F_i(t)$ which tracks the beginning and the
end of the transmission of a packet for a flow. These are defined as,
\begin{equation}
S_i^k(t) = max( V(a^k_i), F_i^{k-1} ), \>\>\>\>\>  F_i^0 = 0 
\label{VST:eqn}
\end{equation}
\begin{equation}
F_i^k(t) = S_i^k + \frac{L_i^k}{\phi_i}
\label{VFT:eqn}
\end{equation}
where $S_i^k$ and $F_i^k$ denote the virtual start time and virtual finish time 
respectively for the $k^{th}$ packet of session $i$, $L_i^k$ is the length of the 
$k^{th}$ packet of session $i$, and $a^k_i$ is the arrival time of the $k^{th}$ packet 
of session $i$. We will see that the first operation can be of complexity O(N) to O(1),
depending on how closely it emulates FFQ. The second operation is a sorting operation 
and the best we can do is $O(logN)$. 

{\em Weighted Fair Queueing} (WFQ) \cite{pgps} simulates the FFQ server to track the 
system virtual time. The system virtual time is updated every time a packet departs. 
FFQ server can service $N$ flows within this time. Hence update of the system virtual 
time in the worst case is an O(N) operation. On arrival of a packet to a flow, it is 
stamped with the virtual finish time, which is used to insert it into a priority queue 
for dispatch. This takes $O(logN)$ time. WFQ gives tight delay bound compared to FFQ. It 
has been shown that the maximum delay for a packet in WFQ is always within one packet 
transmission time of that of FFQ. But WFQ is unsuitable for implementing in a fast 
switch because when millions of flows are present the O(N) time complexity becomes a 
bottleneck.

Other approaches have tried to break the O(N) bottleneck by choosing different ways
to calculate the finish time at the cost of loosening the delay bounds. In 
{\em Self-Clocked Fair Queueing} (SCFQ) \cite{scfq}, Golestani uses the finish time of 
the packet being serviced for estimating the system virtual time. In contrast, 
{\em Start-Time Fair Queueing} (SFQ) \cite{sfq} uses the start time tag on each packet 
as the priority index to decide the departure schedule.  A combined approach is used in 
{\em Minimum Starting-Tag Fair Queueing} (MSFQ) \cite{msfq} where packets are sent out 
in increasing order of their finish times, but system virtual time is updated to the 
minimum of the start time of all the backlogged flows. The complexity for all these 
approaches is still $O(logN)$.

\mycomment{
RPS builds methodology to design scheduling algos with same end-to-end delay bound as 
WFQ and bounded fairness, but reducing the complexity. They define some system
potential function and gives the constraints on it to enable the condition. 
}
Stiliadis and Varma presented a methodology for designing scheduling algorithms that 
provide the same end-to-end delay bound as that of WFQ and bounded fairness, but reduces
the complexity of the algorithms \cite{rps}. Similar to the system virtual time, the 
Rate Proportional Server (RPS) methodology defines the concept of {\em system potential} 
$P(t)$ at time $t$, which keeps track of the state of the system. Each connection is 
associated with a {\em connection potential}, which keeps track of the normalized service 
received by the connection during a busy period, plus any normalized service it missed 
during the period it was idle. The definition of RPS only specifies the constraints on the 
system potential, but no exact method is specified for maintaining it. This helps it to 
encompass different algorithms in this class. The constraints on $P(t)$ are, (i) in a busy 
period the system potential must increase to match the increase in the normalized service 
given to the backlogged queue, and (ii) $P(t)$ is always less than or equal to the 
potential of all backlogged sessions at any time $t$. Having said this, it should be 
noted that RPS class of algorithms are still bounded by the $O(logN)$ time complexity.

\mycomment{
A general framework for characterizing these algorithms were presented by 
Stiliadis and Varma as the {\em Rate Proportional Servers} (RPS) \cite{rps}.
The idea was to give an analytical outline for designing scheduling disciplines
that will have the delay bounds of Fair Queueing servers, can provide bounded
fairness properties and have a simple timestamp computation. The schedules in
the RPS class uses the concept of a {\em system potential}, which maintains
the global state of the system. For each connection there is a {\em per-session
potential}. The main constraint on the {\em system potential} is that it must 
be minimum of the potential of the backlogged sessions. This is because the 
potential of a newly backlogged session is set to the system potential, and 
to achieve a zero latency server requires the above condition. The 
{\em per-session potential} keeps track of the normalized service actually 
received by the connection during an active period, plus any normalized service 
it missed during the period it was not backlogged.  
% I had to skip FFQ and SPFQ for the time being ...
}

So far we have seen the scheduling algorithms that are based on FFQ. They present a 
constant delay bound, but their runtime complexity is bounded by O(logN). Now we will
look at the other class of schedulers based on {\bf Round Robin} scheduling. 
It maintains priority classes corresponding to different quality of service guarantees, 
and traffic is split based on their reservations into one of these priority classes. 
This gives a class-based guarantee (not per-flow) and may starve low priority traffic. 
The advantage is that scheduling complexity becomes O(1), but the delay bound is of the 
order of the number of flows, i.e. O(N).  

{\em Weighted Round-Robin} assigns a weight proportional to the bandwidth 
requirement of the flow to each queue. The scheduler sends packets from the queue till
its time share expires. The Deficit Round Robin (DRR) scheduling, proposed by Varghese 
et al. \cite{drr} extends the idea to include a credit based round robin. A queue 
which has a reservation but is idle at present can accumulate credits for the future 
interval. So in the long-term the service received by the flow is proportional to its 
reservation. However lack of short-term fairness implies bursty traffic. {\em Smoothed
Round Robin} \cite{srr} proposed by Guo gives short-term fairness and better schedule 
delay bound than DRR, besides keeping the O(1) complexity. The cause for burstiness in
DRR is that multiple packets may be sent from one flow to fulfill the bandwidth 
guarantee. Instead in SRR, only one packet (packets are assumed equal-sized) from each
flow is dispatched at a time, and in one {\em round} the flow gets to send $k$ times, 
where $k$ is proportional to its reservation. Assuming 4 flows, the schedule may look 
like \{f3, f2, f3, f4, f3, f2, f3, f1, f4, f3, f2, f3, f4, f3, f2, f3\}, where the
weights for f1, f2, f3, and f4 are 1, 4, 8, 3 respectively. In a similar earlier work
\cite{dfq}, Prashant Pradhan had presented a scheme called {\em Discretized Fair 
Queueing} (DFQ). In his formulation, he chooses a granularity parameter $T$ and builds 
a round by selecting from each flow $(w_i * T)$ bits to send, where $w_i$ is the weight 
of the $i^{th}$ flow. DFQ explains how to implement such a scheduler because it may 
involve fragmenting a packet into multiple $(w_i * T)$ sized chunks. It shows that for 
any two flows with parameter T, over any interval of time when both flows are 
backlogged, the normalized service received is bounded by $2T$. DFQ seems to present a 
more practical setting for a SRR like scheduling algorithm.          

We have pointed out in Section \ref{switch-sched:sec} the limitation of OQ-ing with 
respect to memory bandwidth requirement. CIOQ switches presented a suitable alternative.
Stephen and Zhang designed a scheduling algorithm for CIOQ switches with QoS guarantees.
The {\em Distributed Packet Fair Queueing} (D-PFQ) \cite{dpfq} algorithm places Fair
Queueing Servers at the three contention points: inputs, outputs and the crossbar 
because to meet a packet's deadline it must be fairly scheduled at all these points.
This setup requires ($N^2 + 3N$)\footnote{$N^2$ servers at inputs, $N$ servers each at 
ingress and egress to the crossbar, $N$ servers at output.} PFQ servers. In practice we 
need $N+1$ servers, as the $N$ servers at one input and the server at the ingress to the 
crossbar can share the same PFQ server. Now packets destined to an output $j$ is
arriving from all the $N$ input ports. Therefore, $N$ PFQ servers at the input and one 
server at the egress of the crossbar, i.e. $N+1$ servers must be coordinated to 
emulate a single PFQ server at the output of an OQ switch. The difficulty is each of the
$N+1$ servers have separate system virtual times, but we need to normalize these $N+1$ 
different times to compare them. This is done by maintaining an offset value at the 
entry and exit to the switch core, which is used to calculate the normalized timestamp. 

The complexity of managing $N+1$ servers can be avoided if somehow it is possible to
generate the order in which the packets are going to depart from an output queued switch.
This was addressed by Neogi in the Physically Input Queued and Logically Output Queued 
(PILO) architecture \cite{pilo}. In the PILO architecture, for every incoming packet the
arrival time and the flow identifier is transferred to the output, while the actual
packet is held at the input buffer. Using the metadata the output, which is running a
fair queueing server, can generate the departure order. The centralized switch scheduling 
algorithm uses the order at the outputs to select the matching between the inputs and the 
outputs during each time slot.

%###################################################################################

\input{proposal.tex}

%###################################################################################

\Se{ARCHITECTURES FOR NETWORK PROCESSOR}\label{architecture:sec}
The design of a Network Processor features the use of a number architectural techniques
to speed up the processing. In this section we will look at some of the commonly used
techniques and compare their relative merits and demerits. There are a number of 
manufacturers of NPs in the market. We will look at some of the innovative design 
choices made by few of the leading companies.

\Sse{Network Processor Design Space}
\mycomment{
* Pipelining with the use of ASICs.
* Parallelism or Multi Processors
* Fine Grained Multithreading
* Simultaneous Multithreading
* Caching
}
The use of Application Specific ICs (ASICs) had been the most common way to speed up
processing in a router. Though NPs are looked upon as replacement to the ASICs, it still
makes use of ASICs to speed up specialized functions. Many NPs use the idea of
{\bf pipelining}. They break up the packet processing task into a number of stages, like
parsing, classification, queueing and management/modification of packets. Specialized 
co-processors, which form the functional units for each stage of the pipeline, are 
used to implement these functions.

The use of {\bf Superscalar} (SS) processor architecture can also speed up packet 
processing by issuing multiple instructions simultaneously. If packet processing tasks
have sufficient instruction level parallelism (ILP), SS architecture can exploit the ILP.
Further speed up is possible by doing {\bf Fine-Grain Multithreading} (FGMT). In an
FGMT architecture, instructions may be fetched and issued from a different thread of 
execution each cycle. The FGMT architecture attempts to find and exploit all available 
ILP within a thread of execution, but can also mitigate the latency due to a stalled thread 
by quickly switching to another thread. This can improve the overall system throughput.  

Since at most one thread can be executed at a time, the above architectures cannot boost
the performance if the parallelism is among the threads of execution. {\bf Simultaneous
Multithreading} (SMT) \cite{smt} exploits the thread-level parallelism (TLP). A SMT 
architecture has hardware support for multiple thread contexts and extends instruction 
fetch and issue logic to allow instructions to be fetched and issued from multiple 
threads each cycle. Thus, the instruction throughput can be increased according to the
amount of ILP available within each thread and the amount of TLP available between 
threads. A similar way of processing multiple packets independent of another is to use
{\bf Multiple Processors}. If multiple processing cores can be embedded on a single 
chip to build a Chip Multiprocessor (CMP) then substantial throughput gain can be
achieved by permitting multiple threads to execute completely in parallel. Each of the
processors on a CMP will have separate execution pipelines, register files, fetch 
units, etc. CMP has the limitation that a thread may use only the resources bound to
the processor it is executing on. With limited chip area, if each processor is given
a single functional unit, it means that the issue width must be 1, preventing the 
advantages from ILP.  

A time-tested trick for speeding up processing is the use of {\bf Cache}. The benefits
of caching in network-centric processing has been shown by Pradhan et al. for doing
fast address lookup \cite{cache-ecsl}. The simplest idea is to cache the result of a 
lookup. When the next packet arrives, first look in the cache for a hit on the 
address in this packet to retrieve the outgoing port. This can save a lot of time because
as we have seen in Section \ref{classification:sec} the forwarding tables are usually
large and are stored in slower DRAMs. The complexity of a Longest Prefix Match 
algorithm also limits the performance. The naive caching however fails to achieve 
expected gains because the base assumption of temporal correlation in network traffic 
behavior cannot be the same as program behavior. There is noticeable improvement if we 
store host address ranges, instead of host addresses. This is called Host Address Range 
Caching (HARC). Now the number of outcomes for route lookup is bounded by the number of 
output interfaces regardless of the size of the routing table. This observation leads to an 
algorithm to select the index bits into the cache in such a way that ranges with the same
output are merged. This minimizes the number of distinct ranges and better hit ratio can 
be obtained because capacity misses are reduced. Gopalan et al. extended the algorithm
to reduce the conflict misses by choosing the index bits such that ranges are mapped 
uniformly to cache sets, thus reducing the hot-spots in cache sets \cite{cache-ecsl2}.

\Sse{Commercial Network Processors}
\mycomment{
Major NP manufacturers :
	INTEL, 
	IBM, 
	Lucent Agere, 
	MMC Networks.
}
In this subsection, we will look at some of the industry-standard Network Processors. 
The wide differences among the NPs stem from a number of factors. Firstly, different
NP vendors target different router market segments, which consist of backbone routers,
edge routers and access routers. The requirements are different in each of these cases 
in terms of line speeds and number of ports supported, how deep in the OSI layer the 
processing need is, and how complex the processing is with regard to data and control 
plane. For example, we can see that there are NPs with support for 16 Fast Ethernet (100
Mbps) links to a single OC-192 (10 Gbps) link. Secondly, the core architecture differs 
with respect to the number of processing elements used, the use of specialized hardware,
on-chip bandwidth, as well as memory requirements. Thirdly, the NPs are not stand-alone
systems. Hence they need to have well-defined interfaces to other system components like
the off-chip memory, the switch fabric, the physical media and the host processor. As an
example, the media interfaces can be Fast Ethernet, Gigabit Ethernet, OC-3 POS, 
OC-12 POS, OC-12 ATM, OC-48 POS and/or OC-192 POS. We will take a look at the NPs along
these dimensions.     

%###################################################################################
\myfig{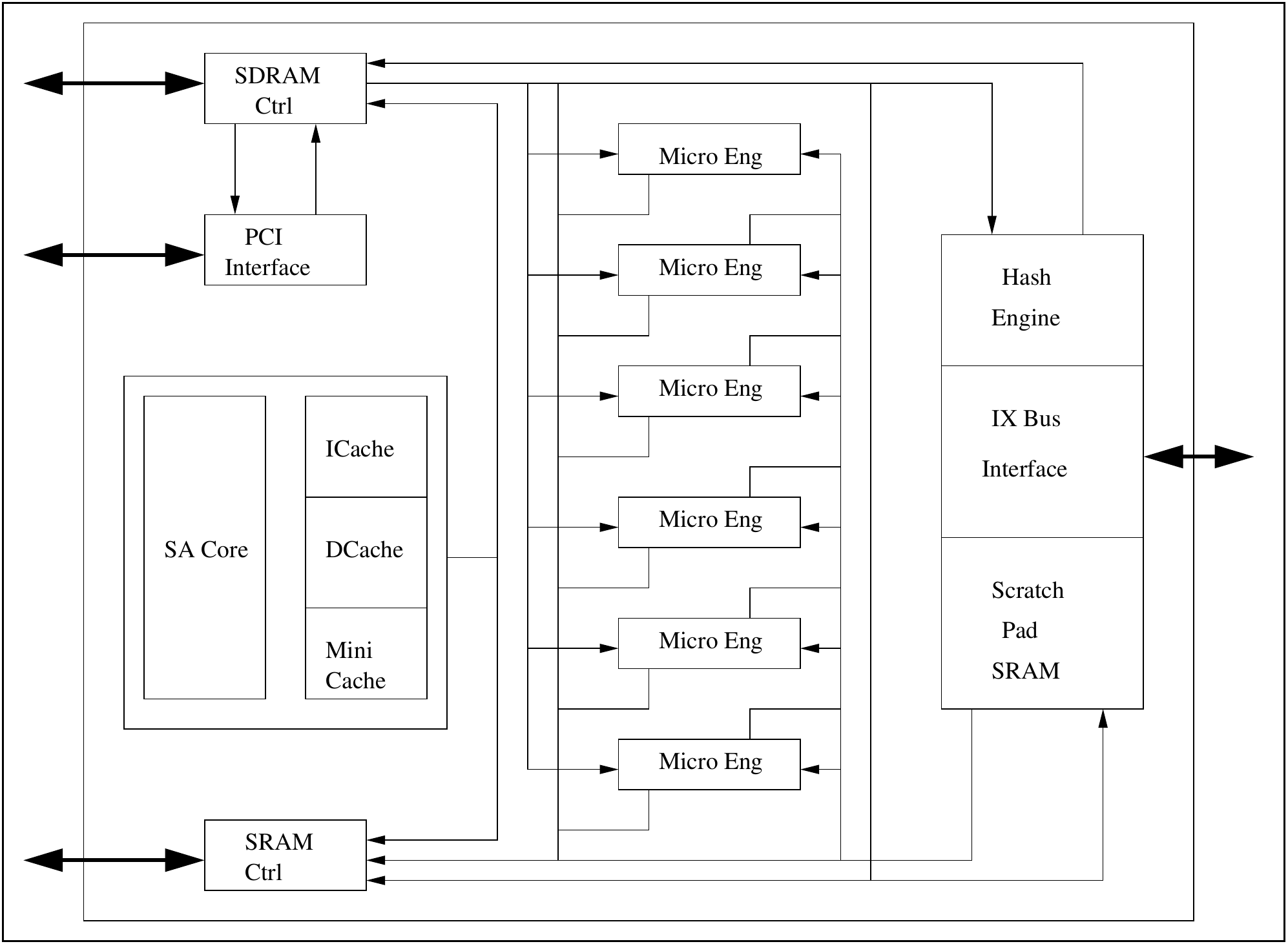}{4.0in}{\sl Intel's IXP1200 Architecture.}
%###################################################################################

{\bf Intel} is one of the leading manufacturers of NPs with a line of NPs named
IXP1200 \cite{ixp1200}, IXP1240, IXP1250. The IXP1200 class of NPs are aimed at
layer 2-4 processing and can support a packet rate of 2.5 Mpackets/s. It consists of six
programmable micro-engines and a 200 MHz StrongARM core that coordinates the activities.
The 64 bit wide Intel IX bus connects the micro-engines, StrongARM, memory and the 
off-chip devices. A PCI bus interface allows integration with an external control 
processor. The micro-engines have hardware support 4 threads each, giving a total of 24 
threads on the chip. There are also some specialized hardware for tasks, like hashing.
A block diagram of the IXP1200 is shown in Figure \ref{ixp1200:fig}. Thus Intel's 
approach is similar to doing chip level multiprocessing with thread Multithreading 
support.

{\bf IBM}'s line of NPs called PowerNP \cite{powernp} is also targeted toward layer 
2-5 processing. This is a multi-processor solution with 16 protocol processors, 7 
specialized co-processors, and a PowerPC core. It supports Packets-over-Sonet (POS) 
and Gigabit Ethernet at 2.5 Gbps. Each protocol processor has a 3-stage pipeline. Each 
pair of protocol processors also share hardware co-processor to speed up tree searching 
and frame manipulation. The 7 specialized co-processors are used to accelerate tasks like 
queueing, checksumming, etc. 

{\bf Lucent-Agere's PayloadPlus} \cite{payloadplus} is another noteworthy name in the
NP arena. This is also aimed at layer 2-4 processing at 2.5 Gbps. PayloadPlus comprises
2 specialized processing elements on the data path, called the Fast Pattern Processor 
(FPP) and the Routing Switch Processor (RSP). The FPP has an interface to the physical 
media. An incoming packet first comes to the FPP, which performs the pattern matching, 
and forwards it to the RSP. The RSP takes care of the traffic shaping, queueing, and 
packet modification functions, and interfaces with the switch fabric. The FPP is a 
pipelined, multithreaded processor supporting up to 64 threads. The PayloadPlus also 
provides another functional block called the Agere System Interface (ASI) whose
functions are to manage routing table updates, gather statistics, etc. These functions
are similar to control path processing. Hence the ASI is kept separate from the faster
data path comprising FPP and RSP. The ASI also provides the 64 bit, 66 MHz PCI interface 
with DMA and interrupt support that can be used to connect to a host CPU. It also has
2 32-bit SSRAM memory interfaces.  
 
{\bf MMC Network}s', presently Applied Micro Circuits Corporation (AMCC), nP7xxx family 
of NPs \cite{mmcNP} are aimed at full 7-layer processing. The nP7510 in particular can 
support 1 OC-192 (10 Gbps), or 4 OC-48, or 10 Gigabit Ethernet ports. It uses six 
EPIF-200 packet processors on a single chip. The EPIF-200 is a 64 bit processor with 
network optimized instruction set (NISC) and zero-overhead task switching among 8 
threads. It has a programmable policy engine for packet classification and a search 
engine for layer 2 VLAN bridging and layer 3 Longest Prefix match. Besides it has 
statistics gathering engine that collects RMON-compliant data.  
 
Other players in the market are {\bf EZchip} \cite{ezchip} and {\bf Motorola} 
\cite{cport}. EZchip's NP-1 processor also claims to perform Layer 2-7 processing at 
10 Gbps. NP-1 is dedicated toward data plane processing and a separate interface is 
provided for a control processor. The NP-1 is designed with a number of specialized 
Task Optimized Processors (TOPs). The key idea in NP-1 design is to split the four 
basic packet processing tasks (parse, search, resolve, modify) across small, fast 
processor cores. Optimization of each task processing is done by using a customized 
instruction set for the processor. This is called the Task Optimized Processing Core 
(TOPCore) technology and based on it EZchip has TOPParse, TOPSearch, TOPResolve and 
TOPModify. For increased processing power, the TOPCores are combined to form a 
superpipelined and superscalar architecture. The packet processing is pipelined passing 
packets from TOPParse to TOPsearch to TOPresolve to TOPmodify. The speedup derived 
from pipelining is further boosted by having multiple instruction pipelines to execute 
several different instructions concurrently during a single cycle. To ensure that memory
does not become a bottleneck, all memory needs of the network-specific processor are 
embedded on-chip.

Motorola's C-Port family of Network Processors also claim 7-Layer processing ability 
at 2.5 Gbps speed. It is a single chip multi-processor consisting of 16 channel processors
with 5 co-processors dedicated for tasks like, coordination with external processors, 
table lookup and update, queue management, and fast memory management. The internal 
buses provide an aggregate bandwidth of 60 Gbps. The external interface for Layer 2/3 
connection is UTOPIA and there is PCI bus interface support.

\mycomment{
{\bf ClearWater Network}'s CNP810SP \cite{xstreamlogic} network processor deserves a 
mention for its architecture which uses Simultaneous Multithreading. It is targeted at 
Layers 4-7 for edge devices at 10Gbps. It supports 8 simultaneous threads. Another
interesting architecture aimed at 40 Gbps is presented by {\bf ClearSpeed}. Their
architecture comprises multiple multithreaded array processors (MTAP) and shared 
co-processors connected by a high bandwidth bus. A MTAP consists of multiple 8-bit 
Processing Elements that all execute the same code.  
}
%###################################################################################
\begin{table*}
\centering
\begin{tabular}[h]{||l|c|c|c|c||} \Hl\Hl
{\bf NP Type} & {\bf \# of Processing} & {\bf \# of Pipeline} & {\bf \# of} & 
{\bf Issue} \\ 
  & {\bf Elements} & {\bf Stages}  & {\bf Threads} & {\bf Width} \\ \Hl\Hl
Intel IXP1200 \cite{ixp1200} & 6+1 & - & 4 /p.e. & 1 \\ \Hl
IBM PowerNP \cite{powernp} & 16+7+1 & 3 /p.e. & 2 /p.e. & 1 \\ \Hl
Lucent-Agere \cite{payloadplus} & 2 & - & 64 & 1 \\ \Hl
MMC Networks \cite{mmcNP} & 6 & & 8 & 1  \\ \Hl
EZchip \cite{ezchip} & 4 & 4 & - & - \\ \Hl
Motorola C-Port \cite{cport} & 16+5+1 & - & - & - \\ \Hl
\end{tabular}
\caption{\sl A relative picture of industry standard Network Processors.}
\label{np:tab}
\end{table*}
%###################################################################################
In Table \ref{np:tab} we distinguish these Network Processor
Architectures on the basis of some chosen metrics, like number of processors
used on a chip, number of pipeline stages, number of threads supported, the issue width
for a superscalar processor, and the targeted processing speed. The sources for most of 
the information have been the white papers from the companies and information related 
to all the metrics were not available. In summary, the main target for most of the NPs 
seem to be OC-48c (2.5 Gbps) and OC-192c (10 Gbps).

\Se{BENCHMARKING THE NETWORK PROCESSORS}\label{benchmark:sec}

\mycomment{
* what is BM supposed to tell us ?
* At what granularity do we need the BM: system, application or task. 
	refer NPF BWG
* Can we use the existing BMs, like SPEC ? If not, why ?
* What main features must be present in a NP BM ? ... ( BM Methodology paper )
* Examples of existing BM ... [ Commbench, Mediabench and Netbench ] 
        why do you think they can improve on SPEC ?
}
We saw the diversity of choices in Network Processor design and its reflection in 
the industry standard NPs. The NPs have widely disparate micro-architectures, memory 
architectures, and system-level interfaces (control, physical media and switch fabric). 
This makes evaluation and comparison of NPs a tedious effort. Benchmark suites 
are intended to provide the normalizing platform on which NPs can be compared. In 
this section, we will take a look at the approaches being advocated for choosing
benchmarks to evaluate NPs. There are no standard benchmark available in the community
for NPs. We will look at the academic research and the efforts of the Network Processing 
Forum (NPF) toward this end. 

The Network Processing Forum Benchmarking Working Group (NPFBWG) \cite{npfbwg} advocates
benchmarking efforts at three levels of granularity: system-level, 
application-level and task-level. Benchmark suites at these three levels have a 
hierarchical relationship. The lowest level, which is the task-level benchmarks 
characterizes how rapidly a NP performs common networking tasks that comprise networking
applications, for example Longest Prefix Match lookup. The application-level benchmarks 
are supposed to characterize performance of well-defined application functions, like IP 
forwarding, MPLS switching. System-level benchmarks are supposed to capture the complete
system performance for typical NP application domains, like firewalls. Mel Tsai et al.
provides an even more comprehensive methodology to design benchmarks for NPs 
\cite{ucbBM}. According to them, for each benchmark application separate and precise 
specification of {\em functionality}, {\em environment}, and {\em measurement} must be
provided. Benchmark functionality captures the important aspect of the benchmark's 
algorithmic core. The environment provides the NP features that allows results to be
compared across NPs. The guidelines for measurement of performance give consistency of
results. 

The traditional benchmark suites have been SPEC, Dhrystone. However, in light of the 
discussion, these do not seem to be suitable choice. Firstly, they focus solely on the
functional characteristics of the applications, and cannot account for the differences
arising out of varying architectural platform. Secondly, there is a difference between 
a NP workload compared to the general workload as we will see. The {\em CommBench} suite
\cite{commbench} chooses two kinds of benchmarks, header processing applications (HPA),
and payload processing applications (PPA). The HPA are Radix-Tree Routing Table lookup, 
IP Fragmentation involving computing checksum, Deficit Round-Robin Scheduling, TCP 
monitoring (traffic policing) using {\tt tcpdump}, and the PPA are CAST, which is an
encryption algorithm, ZIP used for data compression, REED used for FEC and
JPEG which is a lossy compression algorithm. This selection fits the category of 
application-level benchmarks, but fail to incorporate a way to define system-level 
interfaces. Commbench shows difference with SPEC along four dimensions: code size,
computational complexity, instruction set characteristics, and cache performance.
The average code size for Commbench is 5750 compared to 48,700 for SPEC. The average
difference in frequencies of executed instructions between Commbench and SPEC was 5\%.
Also, the cache miss rates were different for instruction and data caches for CommBench 
and SPEC. 

Another benchmark suite has been developed by UCLA, called NetBench \cite{netbench}.
They also come up with benchmarks that can be seen as belonging to task-level, 
application-level, and system-level. At task-level we have CRC-32 checksum computation 
and Table lookup in a radix-tree routing table, application-level comprises IPv4 routing,
Deficit Round Robin, Firewall and Network Address Translation (NAT), and at system-level
there are SSL and URL-based switching. 

All the existing benchmark suites lack in providing the common platform to compare NPs.
NPF provides the mandatory information that should accompany the result of any test. It
also provides a list of application-level and task-level benchmarks it plans to specify
\cite{npfbwg}.

\Se{CONCLUSION}\label{summary:sec}

The Internet is marked by two distinct trends: the increasing link speeds and the 
emergence of new applications with strict bounds on service guarantees. The first
development pushes the routers to carry out its data plane processing faster, while
the second requires more processing cycles at each intermediate router. The 
conflicting goals have opened the market for better software and hardware solutions 
to prevent routers from becoming the bottleneck in communication. 

In this survey, we have discussed four main tasks that a router needs to perform 
on every packet traversing it: packet classification, buffer memory management, switch
fabric scheduling and output link scheduling. For each of these operations we have 
compared the algorithms which can boost router performance by breaking the performance 
bounds. In the case of packet classification, despite a number of innovative solutions,
the use of Ternary CAMs proves to be the best approach. However, for routers with large 
forwarding databases TCAMs cannot be used as they come in limited size. We noted that 
none of the algorithms successfully address the space complexity in case of 
multi-dimensional classification. The lowest bound offered is $O(N)$, where $N$ is the 
number of rules in the forwarding database. In case of buffer memory management, 
input queueing is favored because of its low memory bandwidth requirement, but output
queueing is better suited for guaranteed QoS. Combined Input Output Queueing (CIOQ)
presents the preferred solution for routers since it can take the benefit of lower memory 
bandwidth requirement along with QoS guarantees. CIOQ also opened the question of 
efficient switch fabric scheduling which can ensure QoS. We have looked at some of the
recent approaches based on randomized algorithms. We have also proposed an extension to 
the well-known iSlip algorithm to make it suitable for providing QoS guarantees. In link 
scheduling, we observe that the priority based algorithms fail to break the $O(logN)$ 
bound. Round robin scheduling at the output can give $O(1)$ execution complexity but at 
the price of poorer delay bound. 

The hardware option in router market is driven by the use of Network Processors (NP). 
The use of ASICs is gradually making way to the programmable network processing units. 
There are various architectural choices starting from pipelining, multithreading to the 
use of multiple processors to speed up the packet processing in a NP. A slew of Network
Processors are occupying the market and are aimed at different router market segments.
Some of them are focused on Layer 2/3 Processing at OC-48c (2.4 Gbps) speeds, while
others are looking at complete 7 Layer processing at OC-192c (10 Gbps) speed. The 
difficulty in choosing one NP over another can only be resolved by the presence of a
benchmarking suite for NPs. The Network Processing Forum (NPF) Benchmarking WorkGroup
is devoted to this task. However, there are a number of academic efforts, like 
CommBench and NetBench which have tried to define the set of benchmark applications
for NP. Due to differences not only in the architecture, but also in the variety
of interfaces used by the NPs, coming up with a standard and full-proof methodology for
comparing NPs seems to be a difficult and still unsettled problem. 

%%%%%%%%%%%%%%%%%%%%%%%%%%%%%%%%%%%%%%%%%%%%%%%%%%%%%%%%%%%%%%%%%%%%%%%%%%%%

\bibliography{rpe}
\bibliographystyle{plain}

%%%%%%%%%%%%%%%%%%%%%%%%%%%%%%%%%%%%%%%%%%%%%%%%%%%%%%%%%%%%%%%%%%%%%%%%%%%%

\end{document}

%% file: headers/mymacros.tex
\newcommand{\Se}[1]{\section{#1}\label{#1}}
\newcommand{\Sse}[1]{\subsection{#1}\label{#1}}

\newcommand{\Bi}{\begin{itemize}}
\newcommand{\Ei}{\end{itemize}}
\newcommand{\Bn}{\begin{enumerate}}
\newcommand{\En}{\end{enumerate}}
\newcommand{\Bd}{\begin{description}}
\newcommand{\Ed}{\end{description}}
\newcommand{\Bq}[1]{\begin{equation}\label{#1}}
\newcommand{\Eq}{\end{equation}}
\newcommand{\Bqn}[1]{\begin{eqnarray}\label{#1}}
\newcommand{\Eqn}{\end{eqnarray}}
\newcommand{\Hl}{\hline}
 
\newcommand{\BTab} { \vspace{0.3cm} \begin{table}[phtb] \centering }
\newcommand{\ETab}[2] {\caption{#2} \label{#1:tab} \end{table} \vspace{0.3cm} }
\newcommand{\BTAB} { \vspace{0.3cm} \begin{table*}[phtb] \centering }
\newcommand{\ETAB}[2] {\caption{#2} \label{#1:tab} \end{table*} \vspace{0.3cm} }

\newcommand{\junk}[1] {}
\newif\ifsizedfigures \sizedfiguresfalse
\newif\ifincludefigures \includefiguresfalse
\newif\iffigurenamesinmargin \figurenamesinmarginfalse
\newcommand{\myfig}[3]
{
\begin{figure}
\centerline{\psfig{figure=./fig/#1.eps,width=#2}}
\caption{#3}
\label{#1:fig}
\end{figure}
}

\newcommand{\mycomment}[1]
{}

%% file: proposal.tex
\Se{Switch Scheduling Algorithm with QoS guarantee for Input-Queued Switch}
\label{qslip:sec}

In Section \ref{switch-sched:sec}, we have seen that input buffered switches are
preferred for their low memory bandwidth requirement. But most of the scheduling
algorithms dealing with Quality of Service (QoS) assume an output buffered switch,
as we have seen in Section \ref{psched:sec}. There have been some efforts, like the
WPIM algorithm \cite{wpim}, which have tried to give a port based bandwidth
guarantee. In this section our goal is to explore the possibility of providing 
QoS guarantees for input buffered switches. We will present an extension to the 
iSlip algorithm \cite{islip} that can provide fairness in terms of sharing the
bandwidth among the flows destined to the same output, and also bound their delay
to $O(N)$.

%###################################################################################
\myfig{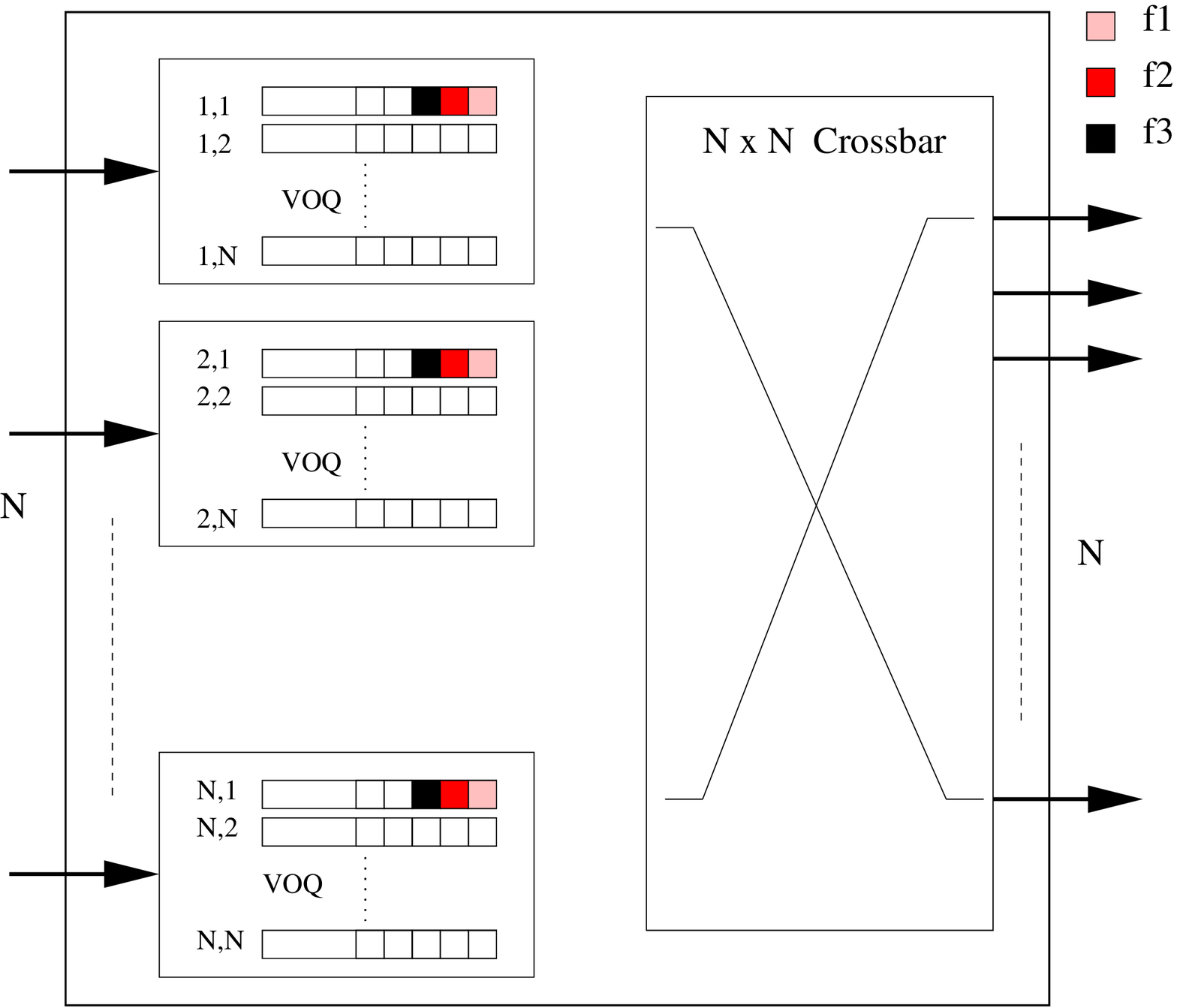}{3.0in}{\small \sl An Input Queued Switch model. The pink, red and black
packets are from three different flows which would be serviced in a Round Robin 
order at the output 1. We will show that an Input Queued Switch can mimic a similar 
behavior.}
%###################################################################################

\Sse{Proposed Algorithm} 
\mycomment{
explain iSlip once more ...
why will iSlip fail to give QoS ...

add a round-robin arbiter at each VOQ ... does it solve the problem ?

add a bitmap of flows pending ... output picks from inputs who has a packet 
to send for the particular flow.
}

The three main steps in the iSlip algorithm are,
\begin{itemize}
\item{\em Request:} Each input sends a request to every output for which it 
has a queued cell.
\item{\em Grant:} If an output receive any requests, it chooses the one that appears 
next in a fixed, round-robin schedule starting from the highest priority element.
The output notifies each input whether or not its request was granted. The pointer
to the highest priority element of the round-robin schedule is incremented to one 
location beyond the granted input if and only if the grant is accepted by an input
in the next step.  
\item{\em Accept:} If an input receives a grant, it accepts the one that appears next
in a fixed, round-robin schedule starting from the highest priority element. The 
pointer to the highest priority element of the round-robin schedule is incremented 
to one location beyond the accepted output.
\end{itemize}

The iSlip algorithm can maximize throughput but it will fail to give a fair share
of the bandwidth to the competing flows. Let us assume that there are 3 flows,
$f1$, $f2$, $f3$, destined to output $1$ and their packets are spread across 
different inputs. In Figure \ref{qslip:fig} we show the IQ switch model we are
using. It has Virtual Output Queues (VOQs) corresponding to each output. The packets
from the 3 flows are spread across the VOQs of all the $N$ inputs. Since, iSlip
uses a rotating priority model, therefore one possible order the packets can be
dispatched to the output is that packet from VOQ(1,1) is sent followed by the packet
from VOQ(2,1) and then the packet from VOQ(3,1). As the packets are dequeued in a 
FCFS order from the VOQs, therefore, the packets from flow $f1$ will be serviced
in 3 successive time slots. We want to ensure that the service order is such that 
packets from $f1$, $f2$, $f3$ are selected successively.

Whenever a packet arrives it is classified to determine its outgoing 
port. We can also determine its flow identifier (flow-id). Hence, if a VOQ has the
knowledge that the last packet it has sent out was from flow $f1$, then
next time when it has to pick a packet for sending it can choose the packet from 
the next flow in the priority order, e.g. $f2$ in the round-robin scheme. This 
implies that maintaining a round-robin arbiter for each VOQ to select the flow can 
be a possible solution. However, this cannot ensure that we strictly follow a 
round-robin ordering at the output. Since, a particular output may give the `grant' 
to different physical inputs, i.e. a different VOQ, in successive time slots, 
therefore the knowledge about which flow's packet was sent in the previous time slot 
is unknown. We assume that no information is shared among the input queues. 

In order to ensure that a packet from the correct flow is sent to the output, the
grant signal from the output can carry the flow-id of the expected flow. If the VOQ 
of the input which received the grant has a packet from that flow buffered in the VOQ,
then it accepts the grant, otherwise it marks this VOQ inactivate for the next 
iteration. However, this scheme might take unbounded amount of time to converge. 
All the VOQs for an output can have a packet to send but none of them may belong to 
the flow that is expected by the output. Hence the request-grant process has to be 
iterated $N$ times before the packet from the next flow can be requested. 

To make the algorithm converge faster the output must make a decision on which input 
to give a grant based on whether it has a packet from the expected flow. We propose
that every VOQ maintain a bitmap which will tell which flows are pending in its VOQ.
The `request' now carries the additional information in the form of this bitmap. The 
output will pick the input which has a packet from the expected flow and is the 
highest in the round-robin priority. If there is no input queue holding a packet
from the particular flow, the output can decide to send a grant for the packet from 
the next flow in the priority. This means that the convergence time for this 
algorithm should remain same as the original iSlip algorithm.